\documentclass[apj,numberedappendix]{emulateapj}
\usepackage{amssymb,amsmath}
\usepackage{graphicx}
\usepackage{appendix}
\usepackage{natbib}
\usepackage[backref,breaklinks,colorlinks,citecolor=blue]{hyperref}
\usepackage{enumerate}
\usepackage{multirow}
\usepackage{bm}

\newcommand{\etal}{et~al.~}

\altaffiltext{\MIT}{Kavli Institute for Astrophysics and Space Research, Massachusetts Institute of Technology, 77 Massachusetts Avenue, Cambridge, MA 02139}

\altaffiltext{\KIPAC}{Kavli Institute for Particle Astrophysics and Cosmology, Stanford University, 452 Lomita Mall, Stanford, CA 94305}
\altaffiltext{\Stanford}{Department of Physics, Stanford University, 382 Via Pueblo Mall, Stanford, CA 94305}
\altaffiltext{\SLAC}{SLAC National Accelerator Laboratory, 2575 Sand Hill Road, Menlo Park, CA 94025}

\altaffiltext{\FNAL}{Fermi National Accelerator Laboratory, Batavia, IL 60510-0500, USA}
\altaffiltext{\KICPChicago}{Kavli Institute for Cosmological Physics, University of Chicago, Chicago, IL, USA 60637}
\altaffiltext{\AAUChicago}{Department of Astronomy and Astrophysics, University of Chicago, Chicago, IL, USA 60637}
\altaffiltext{\ANL}{Argonne National Laboratory, 9700 S. Cass Avenue, Argonne, IL, USA 60439}

\altaffiltext{\Miss}{Department of Physics and Astronomy, University of Missouri, 5110 Rockhill Road, Kansas City, MO 64110}
\altaffiltext{\UChicago}{Department of Physics, University of Chicago, 5640 South Ellis Avenue, Chicago, IL 60637}
\altaffiltext{\CfA}{Harvard-Smithsonian Center for Astrophysics, 60 Garden Street, Cambridge, MA 02138}
\altaffiltext{\UMontreal}{D\'{e}partement de Physique, Universit\'{e} de Montr\'{e}al, C.P. 6128, Succ. Centre-Ville, Montr\'{e}al, Qu\'{e}bec H3C 3J7, Canada}

\altaffiltext{\Huntingdon}{Huntingdon Institute for X-ray Astronomy, LLC}

\altaffiltext{\Princeton}{Department of Astrophysical Sciences, Princeton University, Princeton, NJ 08544, USA}

\altaffiltext{$\dagger$}{Einstein and Spitzer Fellow}

\def\MIT{1}

\def\KIPAC{2}
\def\Stanford{3}
\def\SLAC{4}

\def\FNAL{5}
\def\KICPChicago{6}
\def\AAUChicago{7}
\def\ANL{8}
\def\Miss{9}
\def\UChicago{10}
\def\CfA{11}
\def\UMontreal{12}
\def\Huntingdon{13}
\def\Princeton{14}

\begin{document}


\title{The Remarkable Similarity of Massive Galaxy Clusters From $\lowercase{z}\sim0$ to $\lowercase{z}\sim1.9$}
   
\author{
M.\,McDonald\altaffilmark{\MIT},
S.\,W.\,Allen\altaffilmark{\KIPAC,\Stanford,\SLAC},
M.\,Bayliss\altaffilmark{\MIT},
B.\,A.\,Benson\altaffilmark{\FNAL,\KICPChicago,\AAUChicago},
L.~E.~Bleem\altaffilmark{\KICPChicago,\AAUChicago,\ANL},
M.\,Brodwin\altaffilmark{\Miss},
E.\,Bulbul\altaffilmark{\MIT},
J.\ E.\ Carlstrom\altaffilmark{\KICPChicago,\AAUChicago,\ANL,\UChicago},
W.\ R.\ Forman\altaffilmark{\CfA}
J.\,Hlavacek-Larrondo\altaffilmark{\UMontreal},
G.~P.~Garmire\altaffilmark{\Huntingdon},
M.\,Gaspari\altaffilmark{\Princeton}$^{\dagger}$,
M.\,D.\,Gladders\altaffilmark{\AAUChicago,\KICPChicago},
A.\,B.\,Mantz\altaffilmark{\KIPAC,\Stanford,\SLAC},
S.~S.~Murray\altaffilmark{\CfA}
}

\email{Email: mcdonald@space.mit.edu}   


\begin{abstract}


We present the results of a \emph{Chandra} X-ray survey of the 8 most massive galaxy clusters at $z>1.2$ in the South Pole Telescope 2500 deg$^2$ survey. We combine this sample with previously-published \emph{Chandra} observations of 49 massive X-ray-selected clusters at $0 < z < 0.1$ and 90 SZ-selected clusters at $0.25 < z < 1.2$ to constrain the evolution of the intracluster medium (ICM) over the past $\sim$10 Gyr. We find that the bulk of the ICM has evolved self similarly over the full redshift range probed here, with the ICM density at $r>0.2R_{500}$ scaling like $E(z)^2$. In the centers of clusters ($r\lesssim0.01R_{500}$), we find significant deviations from self similarity ($n_e \propto E(z)^{0.1 \pm 0.5}$), consistent with no redshift dependence. When we isolate clusters with over-dense cores (i.e., cool cores), we find that the average over-density profile has not evolved with redshift -- that is, cool cores have not changed in size, density, or total mass over the past $\sim$9--10 Gyr. We show that the evolving ``cuspiness'' of clusters in the X-ray, reported by several previous studies, can be understood in the context of a cool core with fixed properties embedded in a self similarly-evolving cluster. We find no measurable evolution in the X-ray morphology of massive clusters, seemingly in tension with the rapidly-rising (with redshift) rate of major mergers predicted by cosmological simulations. We show that these two results can be brought into agreement if we assume that the relaxation time after a merger is proportional to the crossing time, since the latter is proportional to $H(z)^{-1}$.
\end{abstract}

%

\section{Introduction}
\setcounter{footnote}{0}

As the most massive collapsed structures in the Universe, galaxy clusters provide unique laboratories for studying physics on very large and energetic scales. In particular, X-ray observations of galaxy clusters, which probe the hot ($\gtrsim$10$^7$K) intracluster medium (ICM), lead to an understanding of cluster-cluster mergers, the most energetic phenomena in the Universe \citep[e.g.,][]{markevitch02,sarazin02}, allow detailed studies of the effects of active galactic nuclei (AGN) on large scales \citep[see reviews by][]{fabian12,mcnamara12}, and provide some of the tightest constraints on the amount and distribution of matter in our Universe \citep[e.g.,][]{mantz10,dehaan16}. The cores of galaxy clusters represent one of the least understood regimes outside of our galaxy  \citep[see review by][]{kravtsov12}, with runaway cooling of the hot ICM \citep[e.g.,][]{fabian94, mcdonald12c} being seemingly held in check by frequent outbursts of AGN feedback \citep[e.g.,][]{rafferty08,hlavacek15} -- a phenomenon that simulations are only recently beginning to reproduce \citep[e.g.,][]{gaspari11, gaspari16}.

While the detailed physics of the ICM in nearby clusters has been studied in depth, the \emph{evolution} of the ICM has only recently become an active area of research. This change is due, in large part, to the success of Sunyaev-Zel'dovich \citep[SZ;][]{sunyaev72} surveys, which select galaxy clusters via their imprint on the cosmic microwave background (CMB) -- an effect that is, in principle, independent of redshift. Since the first discovery of a galaxy cluster via the SZ effect \citep{staniszewski09}, the number of new, distant, SZ-selected galaxy clusters has, on average, more than doubled every year \citep{vanderlinde10, marriage11,planck11,hasselfield13,reichardt13,planck14,bleem15,planck15}. At the same time, optical and near-infrared (NIR) selection (based on galaxy overdensity) has matured, yielding complementary stellar mass-selected galaxy cluster catalogs over similar redshift ranges to the SZ surveys \citep[e.g.,][]{eisenhardt08,muzzin09,brodwin13,rettura14,stanford14}. 


With the rapid growth of NIR- and SZ-selected cluster catalogs has come the ability to study galaxy cluster evolution over an unprecedented range in redshift. However, the majority of the X-ray follow-up of the most distant clusters has focused on single extreme objects, such as XMMXCS J2215.9-1738 at $z=1.46$ \citep{hilton10}, XDCP J0044.0-2033 at $z=1.579$ \citep{tozzi15}, IDCS1426.5+3508 at $z=1.75$ \citep{brodwin16}, and 3C294 at $z=1.786$ \citep{fabian03}. This relative lack of statistically-complete X-ray studies of distant clusters, with few exceptions \citep[e.g.,][]{fassbender11}, is broadly due to the small number of known high-$z$ clusters and the increased exposure times necessary at such high redshifts. Without such samples, our ability to make general conclusions about cluster evolution is severely limited.

In recent years, we have completed a survey of 90 SZ-selected clusters with the \emph{Chandra} X-ray Observatory, spanning $0.25 < z < 1.2$ and with M$_{500} \gtrsim 3 \times10^{14}$ M$_{\odot}$. These clusters were drawn from the South Pole Telescope (SPT) 2500 deg$^2$ survey \citep{bleem15}, and observed to uniform depth with \emph{Chandra} from 2011--2014. These data have advanced our understanding of the evolution of the ICM substantially, allowing detailed evolutionary studies of: ICM cooling in cluster cores \citep{semler12, mcdonald13b}, the average entropy and pressure profiles \citep{mcdonald14c}, AGN feedback \citep{hlavacek15}, ICM metallicity \citep{mcdonald16a}, and ICM morphology \citep{nurgaliev16}, while also providing tight constraints on the amount and distribution of matter in the Universe \citep{bocquet15,chiu16,dehaan16}. These studies benefit from the unique combination of the SPT selection function, which is roughly independent of both redshift \citep[e.g.,][]{bleem15} and the dynamical state of the cluster \citep[e.g.,][]{nurgaliev16,sifon16}, and uniform-depth \emph{Chandra} follow-up, meaning that each cluster was observed for sufficient time to collect $\sim$1500--2000 X-ray photons. The latter allows a consistent analysis over the full redshift range of the sample, free from any biases that are signal-to-noise dependent.

Here we extend those previous studies by including new \emph{Chandra} observations of a mass-selected sample of 8 SPT-selected clusters at $1.2 < z < 1.9$. This represents the first X-ray analysis of a mass-complete cluster sample at $z>1.2$, providing new constraints on the thermodynamic state of massive galaxy clusters only $\sim$1--2 Gyr after their collapse. This epoch is roughly the peak of both star formation \citep[see review by][]{madau14} and AGN activity \citep[e.g.,][]{wolf03}, two processes that can alter the chemical and thermodynamic state of the ICM, respectively. In this work, we focus specifically on properties determined from the X-ray surface brightness, deferring detailed spectroscopic analyses to a future paper. In \S2 we describe the data used in this paper, including the low-$z$ cluster sample from \cite{vikhlinin09a} and intermediate-$z$ sample from \cite{mcdonald13b}. In \S3 we discuss our main results, focusing on ICM density profiles and the X-ray morphology of high-$z$ clusters. In \S4 we place these results in the context of previous works and state-of-the-art simulations, before providing a summary and look towards the future in \S5.

Throughout this work we assume $\Lambda$CDM cosmology with H$_0$ = 70 km s$^{-1}$ Mpc$^{-1}$, $\Omega_M$ = 0.3, $\Omega_{\Lambda}$ = 0.7, and define M$_{500}$ and R$_{500}$ in terms of the critical density: M$_{500} \equiv \frac{4\pi}{3}500\rho_{crit}(z)R_{500}^3$.

\section{Data \& Analysis}

\begin{deluxetable*}{c c c c c c | c c | c c}[h!]
\tablecaption{X-ray Properties of SPT-Hi$z$ Sample}
\tablehead{
\colhead{} &
\colhead{} &
\colhead{} &
\colhead{} &
\colhead{} &
\colhead{} &
\multicolumn{2}{c}{Peak} &
\multicolumn{2}{c}{Centroid}
\\
\colhead{Name} &
\colhead{RA} & 
\colhead{Dec} & 
\colhead{z} &
\colhead{M$_{500}$} &
\colhead{R$_{500}$} &
\colhead{$a_{phot}$} &
\colhead{$n_{e,0}$} &
\colhead{$a_{phot}$} &
\colhead{$n_{e,0}$} 
\\
\colhead{} &
\colhead{[$^{\circ}$]} &
\colhead{[$^{\circ}$]} &
\colhead{} &
\colhead{[10$^{14}$ M$_{\odot}$]} &
\colhead{[Mpc]} &
\colhead{} &
\colhead{[10$^{-2}$ cm$^{-3}$]} &
\colhead{} &
\colhead{[10$^{-2}$ cm$^{-3}$]} \\
}
\startdata
\\
SPT-CLJ0156-5541 &  29.0405 & -55.6976 & 1.281 &  3.90$_{- 0.40}^{+ 0.57}$ &  0.69 &  0.09$_{- 0.06}^{+ 0.25}$ &  0.83$_{- 0.14}^{+ 0.17}$ &  0.09$_{- 0.05}^{+ 0.10}$ &  0.81$_{- 0.06}^{+ 0.06}$\\
SPT-CLJ0205-5829 &  31.4459 & -58.4849 & 1.322 &  3.44$_{- 0.40}^{+ 0.63}$ &  0.65 &  0.73$_{- 0.20}^{+ 0.36}$ &  0.93$_{- 0.27}^{+ 0.37}$ &  0.55$_{- 0.18}^{+ 0.36}$ &  0.60$_{- 0.17}^{+ 0.24}$\\
SPT-CLJ0313-5334 &  48.4813 & -53.5718 & 1.474 &  2.01$_{- 0.31}^{+ 1.54}$ &  0.56 &  0.12$_{- 0.20}^{+ 0.64}$ &  0.75$_{- 0.26}^{+ 0.41}$ &  0.11$_{- 0.21}^{+ 0.38}$ &  0.64$_{- 0.24}^{+ 0.37}$\\
SPT-CLJ0459-4947 &  74.9240 & -49.7823 & 1.85$^{\dagger}$ &  2.40$_{- 0.27}^{+ 0.25}$ &  0.49 &  0.46$_{- 0.09}^{+ 0.07}$ &  4.54$_{- 1.09}^{+ 1.43}$ &  0.51$_{- 0.10}^{+ 0.07}$ &  1.98$_{- 0.19}^{+ 0.21}$\\
SPT-CLJ0607-4448 &  91.8940 & -44.8050 & 1.482 &  2.65$_{- 0.36}^{+ 0.55}$ &  0.56 &  0.07$_{- 0.03}^{+ 0.05}$ &  5.98$_{- 1.27}^{+ 1.61}$ &  0.10$_{- 0.05}^{+ 0.05}$ &  3.81$_{- 1.12}^{+ 1.58}$\\
SPT-CLJ0640-5113 & 100.0720 & -51.2176 & 1.313 &  2.92$_{- 0.24}^{+ 0.61}$ &  0.63 &  0.08$_{- 0.02}^{+ 0.03}$ &  3.03$_{- 0.51}^{+ 0.61}$ &  0.07$_{- 0.02}^{+ 0.03}$ &  3.30$_{- 0.47}^{+ 0.55}$\\
SPT-CLJ2040-4451 & 310.2417 & -44.8620 & 1.478 &  3.10$_{- 0.47}^{+ 0.79}$ &  0.60 &  0.35$_{- 0.12}^{+ 0.22}$ &  1.91$_{- 0.62}^{+ 0.91}$ &  0.36$_{- 0.14}^{+ 0.26}$ &  0.54$_{- 0.16}^{+ 0.22}$\\
SPT-CLJ2341-5724 & 355.3533 & -57.4166 & 1.258 &  3.37$_{- 0.34}^{+ 0.70}$ &  0.67 &  0.28$_{- 0.04}^{+ 0.05}$ &  2.09$_{- 0.29}^{+ 0.34}$ &  0.18$_{- 0.03}^{+ 0.05}$ &  2.70$_{- 0.49}^{+ 0.60}$\\
\enddata
\tablecomments{Properties of the clusters in the SPT-Hi$z$ sample. Unless otherwise noted, quoted redshifts are based on spectroscopy of $\sim$5--10 members per cluster. All 8 of these clusters have deep \emph{Chandra} observations, from which we derive M$_{500}$ based on the M$_{gas}$--M relation from \cite{vikhlinin09a}. We provide a quantitative estimate of the X-ray asymmetry ($a_{phot}$) and the central electron density ($n_e$), measured with reference to the X-ray peak and the large-scale centroid of the X-ray emission, measured in an annulus from 250--500 kpc.\\
$^{\dagger}$: Redshift is derived based on a combination of HST and \emph{Spitzer} red sequences, along with X-ray spectroscopy (see \S2.1.1).}
\label{table:data}
\end{deluxetable*}

\subsection{Samples}
In this work, we attempt to trace the evolution of clusters from $z\sim0$ to $z\sim1.9$. This is done by combining the low-$z$ X-ray-selected sample from \cite{vikhlinin09a} with SPT-selected samples at intermediate- \citep{mcdonald13b} and high-$z$. Where appropriate, we apply a mass cut to the X-ray samples to ensure a clean comparison across all redshifts, as shown in Figure \ref{fig:Mz}. Below we discuss the specific details of each data set, including the origin, availability, and quality of X-ray data.

\subsubsection{SPT-Hiz: $1.2 < z < 1.9$}

\begin{figure}[h!]
\centering
\includegraphics[width=0.49\textwidth]{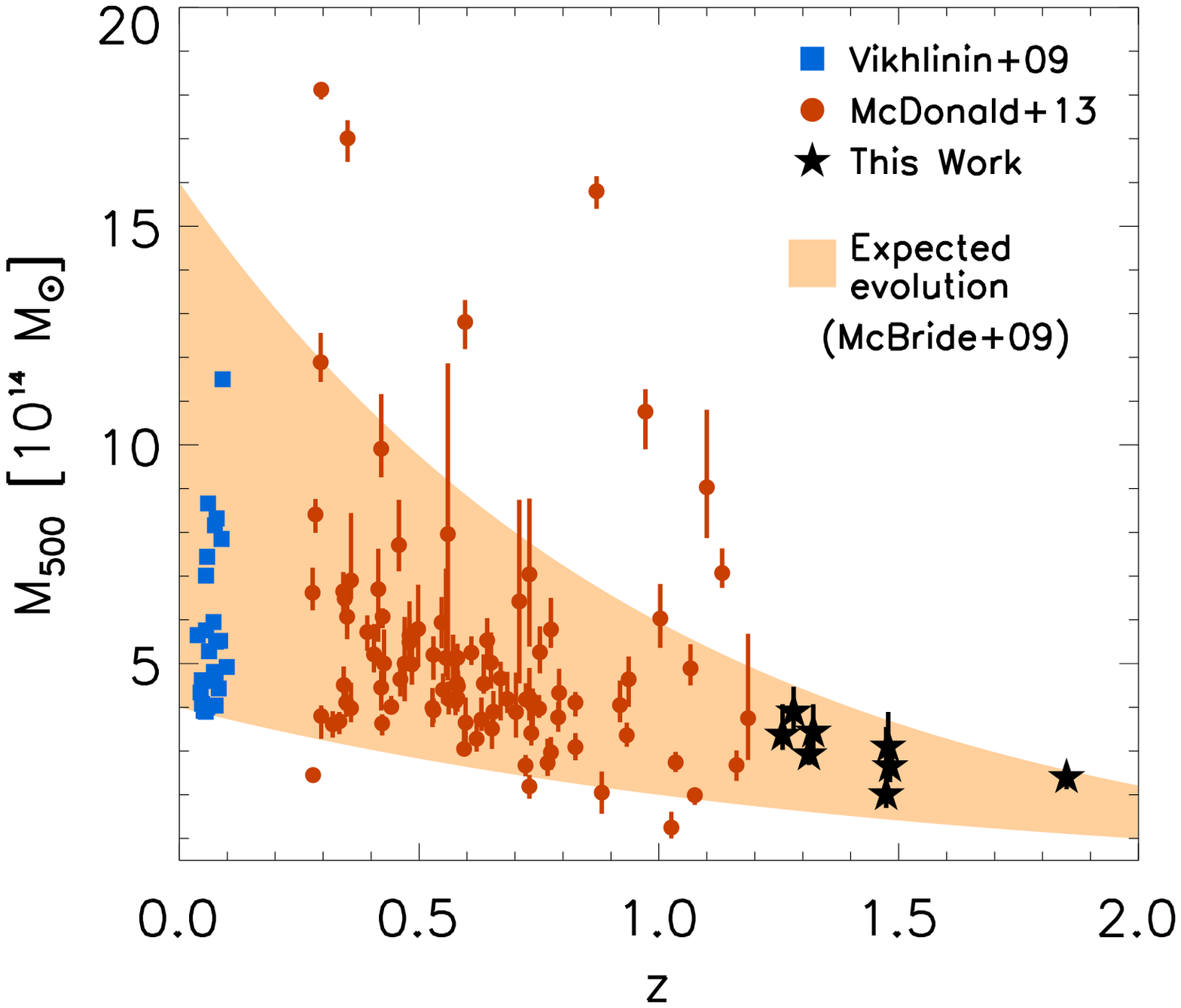}
\caption{Mass versus redshift for the three cluster samples described in \S2.1. The black stars represent the new clusters presented in this work, while the red circles and blue squares show data from \cite{mcdonald13b} and \cite{vikhlinin09a}, respectively. The shaded tan region shows the expected growth track for clusters with M$_{500}$ $\sim$ 2--3 $\times$ 10$^{14}$ M$_{\odot}$ at $z\sim1.5$, from \cite{mcbride09}. This demonstrates that the clusters we are observing at $z>1.2$ are the progenitors of the intermediate- and low-$z$ samples to which we compare.}
\label{fig:Mz}
\end{figure}

The high-$z$ sample, referred to hereafter as ``SPT-Hi$z$'', consists of the 8 most massive galaxy clusters at $z>1.2$ in the 2500 deg$^2$ SPT-SZ survey \citep{bleem15}. These clusters have $2\times10^{14}$ M$_{\odot} <$ M$_{500} < 4\times10^{14}$ M$_{\odot}$ and $1.2 < z < 1.9$, as shown in Figure \ref{fig:Mz}. \emph{Chandra} observations were obtained for each of these clusters as part of a Cycle 16 Large Program (PI: McDonald). For each cluster, we aimed for a total of 1500 counts, where the expected luminosity was derived from the SZ signal assuming the $\xi$--M \citep{bleem15} and the M--L$_X$ \citep{vikhlinin09a} relations. This number of counts has been demonstrated to yield reliable single-temperature and metallicity estimates \citep{mcdonald16a}, allow the measurement of the gas density out to $\sim$R$_{500}$ \citep{mcdonald13b}, and determine accurate X-ray morphologies \citep{nurgaliev13,nurgaliev16}.

Spectroscopic redshifts for most of these clusters are derived based on Low Dispersion Survey Spectrograph \citep[LDSS3;][]{ldss3} spectroscopy of $\sim$5--10 member galaxies per cluster (Bleem \etal in prep), with three exceptions. SPT-CLJ0205-5829 and SPT-CLJ2040-4451, among the earliest clusters confirmed, have optical spectroscopy presented in \cite{stalder13} and \cite{bayliss14}, respectively. SPT-CLJ0459-4947 was not detected in our deep spectroscopic follow-up campaign. However we have deep \emph{Hubble Space Telescope} (HST) imaging of this cluster with WFC3-UVIS and WFC3-IR, which reveals a rich red sequence, allowing us to measure a photometric redshift (Strazzullo \etal in prep). We also have independent redshift constraints for this system from \emph{Spitzer} photometry and from a spectroscopic analysis of the \emph{Chandra} data presented here. Independently, we measure $z=1.85$, $z=1.84$, and $z>1.5$ from the HST, \emph{Chandra}, and \emph{Spitzer} data for SPT-CLJ0459-4947. We adopt a redshift of 1.85 for this system, but stress that the accuracy is at the $\Delta z \sim 0.1$ level. Given that the majority of the analysis presented here requires us to bin all 8 systems at $z>1.2$ into a single average system, the precise redshift of this single system is relatively unimportant.

\subsubsection{SPT-XVP: $0.25 < z < 1.2$}

We include in this analysis a sample of 90 galaxy clusters spanning $0.25 < z < 1.2$ which has been referred to as the ``SPT-XVP'' sample in previous works \citep{mcdonald13b,mcdonald14c}. The bulk of these clusters were observed by \emph{Chandra} via an \emph{X-ray Visionary Program} (hence the name) to obtain shallow X-ray imaging of the 80 most massive SPT-selected clusters at $z>0.3$ (PI: Benson). Additional \emph{Chandra} observations were obtained through various smaller GO (PIs: McDonald, Mohr) and GTO (PIs: Garmire, Murray) programs, or were already available in the archive. For the most part, these observations are of similar depth, with $\sim$2000 X-ray counts per cluster \citep[see Figure 2 in][]{mcdonald14c}. Details of these clusters (selection, masses, redshifts, positions) are provided in \cite{bleem15}, while additional information about the X-ray follow-up can be found in \cite{mcdonald13b,mcdonald14c}. With few exceptions, clusters are selected for X-ray follow-up by mass, with the $\sim$20\% most massive clusters in the full SPT-SZ survey having \emph{Chandra} X-ray observations. The masses and redshifts of these clusters are shown in Figure \ref{fig:Mz}.

\subsubsection{Low Redshift Clusters: $0.0 < z < 0.1$}
For a low-redshift comparison we use the sample of 49 X-ray selected clusters from \cite{vikhlinin09a}. This sample was chosen due to the similarity between our X-ray analysis pipeline and that used in \cite{vikhlinin09a} (the former was modeled after the latter). 
We direct the reader to \cite{voevodkin04} and \cite{vikhlinin09a} for a detailed discussion of how these clusters were selected. In short, the sample is X-ray flux-limited, and constrained in redshift between $0.025 < z < 0.1$. The fraction of merging clusters (defined by eye) in this sample \citep[$31\pm8$\%;][]{vikhlinin09a} is similar to that in the REXCESS sample \citep[$39\pm12$\%;][]{pratt09} and in the SPT-XVP sample (20$^{+7}_{-4}$\%; Nurgaliev \etal 2016).
Each cluster in this low-$z$ sample has deep \emph{Chandra} data, from which we have gas density and temperature profiles from \cite{vikhlinin09a}. From this sample, we only consider clusters with $M_{500} > 4\times10^{14} M_{\odot}$, to allow a fair comparison to the high-$z$ SZ-selected clusters (see Figure \ref{fig:Mz}). This yields a sample of 27 X-ray selected clusters with masses spanning $4\times10^{14} < M_{500} < 1.2\times10^{15}M_{\odot}$. Assuming realistic evolution scenarios for massive halos \citep{mcbride09}, the clusters in the SPT-Hi$z$ sample, which have typical masses of 2--3 $\times$ 10$^{14} M_{\odot}$, will ultimately end up having $M_{500} > 4\times10^{14} M_{\odot}$ at $z\sim0$. 

\subsection{X-ray Data Reduction}
The analysis pipeline used in this analysis was adapted from \cite{vikhlinin06a} and \cite{andersson11}, and is described in detail in \cite{mcdonald13b} and \cite{mcdonald14c}. We repeat relevant aspects here, but direct readers to any of the aforementioned references for additional details.

All \emph{Chandra} data for the SPT-XVP and SPT-Hi$z$ samples were reduced using \textsc{ciao} v4.7 and \textsc{caldb} v4.7.1. Exposures were initially filtered for flares, before applying the latest calibrations and determining the appropriate blank-sky background (epoch-based).  Due to the small angular size of distant clusters, we were able to use off-source regions on the ACIS-I chip opposite the cluster to model the astrophysical background for each observation. In general, these regions were $>$3R$_{500}$ from the cluster center. Blank-sky background spectra were rescaled based on the observed 9.5--12.0 keV flux, and combined with off-source regions to constrain the instrumental, particle, and astrophysical backgrounds. Point sources were identified and masked via an automated wavelet decomposition technique, described in \cite{vikhlinin98}. Cluster centers were chosen in two different ways, which we will consider throughout the text. The ``peak'' center was found by heavily binning and smoothing the image on $\sim$12$^{\prime\prime}$ scales, and then measuring the centroid within 50\,kpc of the peak (to allow sub-pixel accuracy). The ``centroid'' center was found by measuring the centroid within a 250--500\,kpc aperture, following \cite{mcdonald13b}. This definition is less sensitive to core structure (e.g., sloshing) and is a better probe of the center of the large-scale dark matter potential. Unless otherwise noted, all measurements shown are with respect to the ``centroid'' center.

\subsection{X-ray Measurements}

In this work, we focus on measurements derived from the X-ray surface brightness, deferring any spectroscopic analysis \citep[aside from the metallicity evolution study already published by][]{mcdonald16a} to a future paper.
For each cluster, we measure gas density profiles following \cite{vikhlinin06a}, \cite{andersson11}, and \cite{mcdonald13b}, and X-ray morphology following \cite{nurgaliev13} and \cite{nurgaliev16}. Below we briefly describe the relevant features of these analyses.

\subsubsection{Gas Density Profiles}

The surface brightness profile for each cluster is extracted in the energy range 0.7--2.0 keV, in 20 annuli defined as follows:
\begin{equation}
r_{out,i}=(a+bi+ci^2+di^3){R}_{500} ~~~i=1...20~,
\end{equation}
where 
$(a,b,c,d) = (13.779,-8.8148,7.2829,-0.15633)\times10^{-3}$ 
and R$_{500}$ is initially estimated based on the M--T$_X$ relation \citep[see][]{andersson11}. 
This binning scheme is chosen to ensure that the profile is well sampled from core to outskirts, and that the innermost bin is always resolved ($>$1 ACIS-I pixel in radius) for clusters at all redshifts. 
For the cluster with the smallest angular size in our sample (SPT-CLJ0459-4947; $z=1.85$, $R_{500}=494$\,kpc), the innermost bin has $r_{out} = 0.7^{\prime\prime}$, corresponding to $\sim$1.5 \emph{Chandra} ACIS-I pixels in radius, or $\sim$3 pixels in diameter. For all pointings, the cluster center is within 1$^{\prime}$ of the on-axis position, meaning that the innermost bin is roughly the size of (or larger than) the PSF.
Following \cite{vikhlinin06a}, we correct surface brightness profiles for spatial variations in temperature, metallicity, and telescope effective area, assuming a universal temperature profile from \cite{vikhlinin06a}, normalized to the measured $kT_{500}$, and a constant metallicity profile. Calibrated (including k-corrected) surface brightness profiles are expressed as an emission measure integral, $\int n_en_p dl$, where $n_e$ and $n_p$ are the electron and proton densities, respectively. To deproject this into a three-dimensional electron density, we model the calibrated surface brightness profile with a modified beta model:
\begin{equation}
n_en_p = n_0^2\frac{(r/r_c)^{-\alpha}}{(1+r^2/r_c^2)^{3\beta-\alpha/2}}\frac{1}{(1+r^3/r_s^3)^{\epsilon/3}} ,
\label{eq:ne}
\end{equation}
which is projected along the line of sight through the full cluster volume, to match the aforementioned emission measure integral. Here, $n_0$ is the density normalization, and $r_c$ and $r_s$ are scaling radii of the core and extended components, respectively. We estimate the three-dimensional gas density assuming $n_e=Zn_p$ and $\rho_g=m_pn_eA/Z$, where $A=1.397$ and $Z=1.199$ are the average nuclear charge and mass, respectively, for a plasma with 0.3Z$_{\odot}$ metallicity. This assumption of constant, unevolving metallicity is well-motivated by recent work \citep{mcdonald16a}.

Gas masses are derived by integrating $\rho_g(r)$ over the cluster volume. We refine our estimate of M$_{500}$ and $R_{500}$ for each cluster by iteratively satisfying the M$_{gas}$--M$_{500}$ relation from \cite{vikhlinin09a}.

\subsubsection{Morphology}

Following \cite{nurgaliev13} and \cite{nurgaliev16}, we quantify the X-ray morphology using the ``photon asymmetry'' ($a_{phot}$) statistic. This statistic quantifies the amount of asymmetry by comparing the cumulative distribution of X-ray counts as a function of azimuth for a given radial annulus to a uniform distribution, computing a probability that these two distributions are different. Combining these probabilities for multiple radial bins provides an overall probability that the cluster has azimuthally uniform brightness.
This statistic, which is sensitive to azimuthal asymmetry, is complementary to statistics which measure the surface brightness concentration \citep[e.g.,][]{vikhlinin07,santos08}. Importantly, this statistic was shown to be unbiased to the quality of the data used, both in terms of angular resolution and signal-to-noise ratio \citep{nurgaliev13}. This makes it optimal for comparing the morphology of clusters at low and high redshift, where both angular resolution and data quality can vary dramatically.

For each cluster we measure $a_{phot}$, with reference to both the peak and centroid centers (see \S2.2). We report these measurements in Table \ref{table:data} for the SPT-Hi$z$ clusters -- those for the SPT-XVP clusters are reported in \cite{nurgaliev16}. We do not directly compare morphological measurements of high-$z$ clusters to low-$z$, X-ray-selected clusters due to a lack of existing $a_{phot}$ measurements for the latter.


\section{Results}

\subsection{Gas Density Profiles}

\begin{figure*}[htb]
\centering
\includegraphics[width=1.0\textwidth]{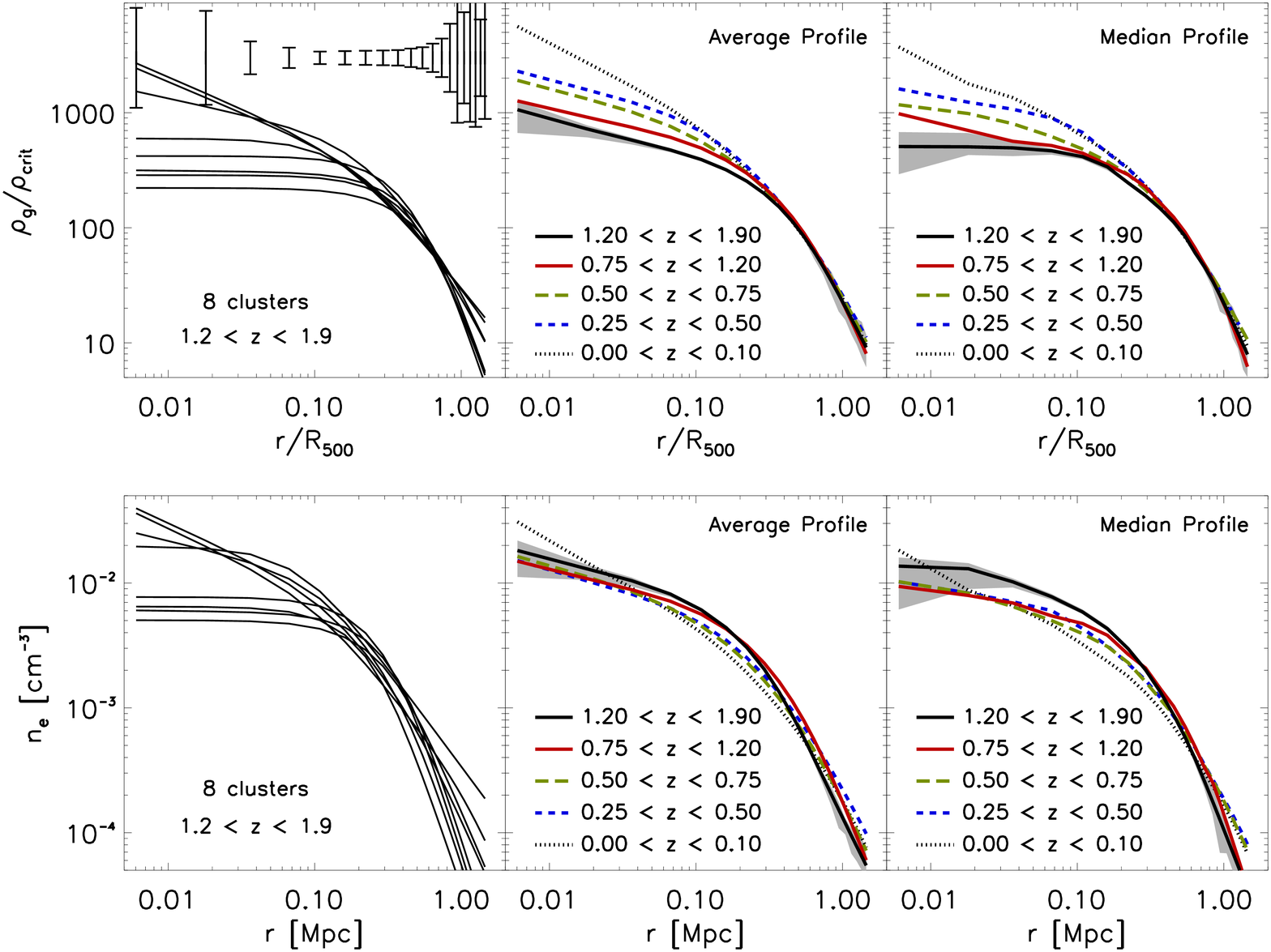}
\caption{\emph{Upper left}: Normalized gas density ($\rho_g/\rho_{crit}$) versus normalized radius ($r/R_{500}$) for the 8 clusters in the SPT-Hi$z$ sample. This panel highlights the large scatter in the cores, where non-gravitational processes such as cooling and feedback can shape the density profile, compared to the small scatter at large radii ($>0.2R_{500}$) where clusters are remarkably self similar. Typical measurement uncertainties in each radial bin are shown at the top, and are dominated by small number statistics at small radii and uncertainty in the background at large radii. \emph{Upper middle}: Average profiles in five different redshift bins. This panel demonstrates that $\rho_g/\rho_{crit}$ in the centers of clusters has increased steadily by a factor of $\sim$5 over the past $\sim$10 Gyr. Outside of the core ($r>0.1R_{500}$), the density profiles appear to be remarkably self similar. The shaded grey band shows the 1$\sigma$ uncertainty in the mean profile for the high-$z$ systems only, for clarity. \emph{Upper right}: Similar to the upper middle panel, but now showing the median profile, rather than the average, which is less sensitive to single extreme systems.  The lack of a measurable cusp in the high-$z$ median implies that the first cool cores may have formed around $z\sim1.6$. The shaded grey band shows the 1$\sigma$ uncertainty in the mean profile for the high-$z$ systems only, for clarity. \emph{Lower panels}: Similar to above, but now showing absolute, rather than normalized, ICM density versus physical radius. These panels demonstrates that much of the ``evolution'' observed in the upper panels may be due to an unevolving central density coupled with an evolving value of $\rho_{crit}$. The scatter in median central ($r<0.012R_{500}$) density over the five redshift bins shown here is only $\sim$10\%.
}
\label{fig:rhog}
\end{figure*}

In \cite{mcdonald13b}, we demonstrated, qualitatively, that the gas density ($\rho_g$) profiles of massive clusters evolve self similarly outside of $\sim$0.15R$_{500}$, over the redshift range $0 < z < 1.2$. In the cores of clusters, this earlier work showed that the ``peakiness'' decreased significantly with increasing redshift, leading to less cuspy density profiles at early times. In Figure \ref{fig:rhog} we extend this earlier analysis to include the 8 SPT-Hi$z$ clusters presented in this work. In the upper left panel of Figure \ref{fig:rhog}, we show the gas density profiles for each of the SPT-Hi$z$ clusters, normalized to the critical density of the Universe ($\rho_{crit} \equiv 3H^2/8\pi G$) and in terms of the scaled radius, $r/R_{500}$. These profiles show an order of magnitude scatter in the innermost bin ($r\sim0.01R_{500}$) and collapse onto a single profile by $r\sim0.3R_{500}$. At large radii, the increased scatter is due to increased noise in the measurements, rather than real, physical scatter as observed in the cores. Next to these individual clusters, we show the average profile in 5 different redshift bins, spanning $0 < z < 1.9$. As in \cite{mcdonald13b}, we see a flattening of the profile with redshift, which appears to extend to $z>1.2$. Given that the average profile can be biased towards cool cores (which have very high central density), we also show the median profile in the right-most panel. The median profile is computed by taking the median density at each radius for all clusters within a given redshift range. This panel demonstrates that the \emph{median} cluster at $1.2 < z < 1.9$ has no visible cusp in the inner density profile ($d\rho_g/dr \sim 0$ for $r<0.1R_{500}$). These data show that, while some clusters at $z\sim1.6$ do have central density cusps \citep[see also][]{brodwin16}, they are in general less peaky than their low-$z$ counterparts.

In the lower panels of Figure \ref{fig:rhog}, we show the electron density profiles in absolute terms, without scaling for the evolving critical density of the Universe ($\rho_{crit}$) or to the evolving (and mass dependent) scale radius ($R_{500}$). These plots highlight what is physically happening to the cluster, and help to clarify the origin of the evolving profiles shown in the upper panels of Figure \ref{fig:rhog}, or in \cite{mcdonald13b}. In the centers of clusters ($r\sim10$\,kpc), \emph{at all redshifts}, the median electron density is $\sim$0.01 cm$^{-3}$, with a measured scatter across 5 redshift bins of only $\sim$10\%. From this common point at the center, the high-$z$ cluster profiles have a shallower inner slope and a steeper outer slope than their low-$z$ counterparts. Likewise, the average profiles have a very small scatter ($<$20\%) in central densities over $0 < z < 1.9$. Given that, over the same redshift range, the critical density of the Universe changes by a factor of $>$5, it is unsurprising that the central values of $\rho_g/\rho_{crit}$ show such a strong evolution (upper panels).

\subsubsection{Deviations from Self Similarity}

\begin{figure}[t]
\centering
\includegraphics[width=0.5\textwidth]{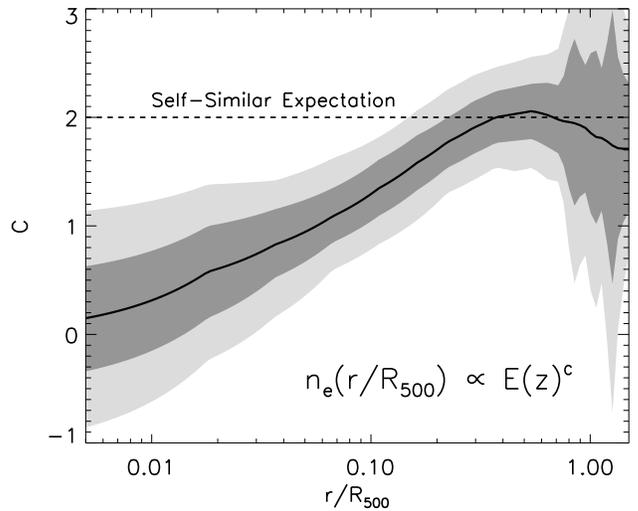}
\caption{Degree to which the radial ICM density profile evolves as a function of redshift. We assume an evolution of $n_e (r/R_{500}) \propto E(z)^C$, with values of $C=0$ and $C=2$ representing no evolution and self similar evolution, respectively. Shaded dark and light regions correspond to 1$\sigma$ and 2$\sigma$ confidence intervals, respectively. This figure demonstrates that, at the centers of clusters, there is no dependence of the gas density on the cluster redshift, while at $r\gtrsim0.2R_{500}$ the evolution is fully consistent with the self similar expectation. This result supports a picture in which the evolution of the core is dictated by local processes (e.g., AGN feedback, stellar feedback, cooling), while the large-scale gas distribution is dictated by gravity.
}
\label{fig:selfsim}
\end{figure}

In the previous section we claim, qualitatively, that the ICM density profile is self similar at large radii, consistent with many previous works \citep[e.g.,][]{vikhlinin06a,croston08,mantz15,mantz16}. Here, we attempt to quantify this degree of self similarity for the full sample of clusters shown in Figure \ref{fig:Mz}. We define 20 radial bins (in terms of $r/R_{500}$; see \S2.3.1), measuring the gas density in each radial bin for each cluster in our sample. We then fit a function of the form $n_e(r/R_{500}) \propto E(z)^C$ within each radial bin, determining the redshift dependence of the density profile at that radius. If the gas density profile evolves self similarly, then it should evolve like $\rho_{crit}$, which scales like $E(z)^2$. In Figure \ref{fig:selfsim} we show how $C$ scales with radius. We find that, at $r\gtrsim0.2R_{500}$, the density profiles are fully consistent (at the 1$\sigma$ level) with self similar evolution ($C=2$). This is consistent with simulations \citep[see e.g.,][]{kravtsov12}, with data from other surveys \citep[see e.g.,][]{mantz16}, and with the general intuition that gravity is the dominant physics at these radii. The large uncertainty in the measurement of $C$ at $r>R_{500}$ is a result of the background emission dominating by a substantial margin at these radii, leading to relatively large systematic uncertainties in the gas density measurement. 

At small radii ($r<0.2R_{500}$), the measured value of $C$ decreases, from $C=2$ at $r=0.2R_{500}$ to $C\sim0$ at $r\sim0.01R_{500}$. This implies a breaking of self similarity in dense cluster cores, where other baryonic physics phenomena (i.e., stellar feedback, AGN feedback, cooling, sloshing, etc) are important. At the centers of clusters, we find no evidence for redshift dependence on the ICM density ($C=0.1 \pm 0.5$), which is akin to the unevolving entropy in cluster cores that we reported in \cite{mcdonald13b}. 
If this result is interpreted as AGN feedback regulating the inner density profile and balancing the multiphase condensation in an inside-out way \citep[e.g.,][]{gaspari14,voit15b}, then it implies that the impact of AGN feedback is confined to $r\lesssim0.2R_{500}$.

While it has long been understood that the density cusps of cool core clusters represent a likely deviation from self similar evolution, we have now directly shown that this is the case using ICM density profiles for clusters spanning $0 < z < 1.9$. We find no evidence that the cores of clusters evolve self similarly, with self similar evolution being ruled out at $>$3$\sigma$ confidence.

\subsubsection{Cool Core Evolution}

\begin{figure}[t]
\centering
\includegraphics[width=0.45\textwidth]{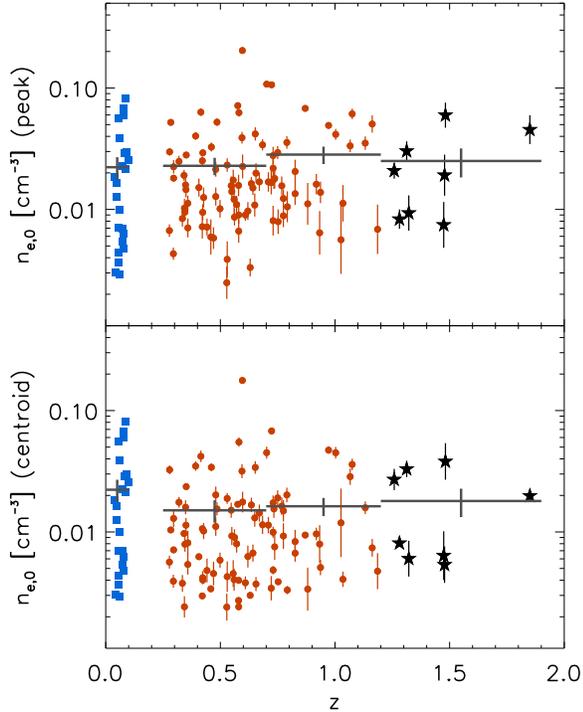}
\caption{Central deprojected ICM density, as measured in the bin $0 < r < 0.012R_{500}$, centered on the peak (upper panel) and centroid (lower panel) of the X-ray emission. Point types and colors are as defined in Figure \ref{fig:Mz}, and correspond to the three different cluster samples used in this work. The large black crosses show the mean and error on the mean for four different redshift bins, demonstrating no measurable evolution in the typical central density of the ICM over $\sim$9.5 Gyr. }
\label{fig:nepeak}
\end{figure}

In Figure \ref{fig:nepeak}, we examine the evolution of the core ICM density more closely, showing the individually-measured central ($r<0.012R_{500}$) densities for all of the clusters considered in this work. For this figure, we define the cluster center in two ways, as described in \S2.2: the peak of the X-ray emission, and the large-scale centroid. We find no measurable evolution in the mean, maximum, or minimum central densities over the full redshift range explored here, independent of the choice of centering method. We note that the centering choice for the clusters from \cite{vikhlinin09a} is slightly different than ours, such that it matches the ``peak'' selection for relaxed clusters, and the ``centroid'' selection for disturbed clusters. As such, it is best compared to the maximum \emph{peak} density, and the minimum \emph{centroid} density. With the exception of the Phoenix cluster at $z=0.597$ \citep{mcdonald12c}, there is a fairly consistent maximum central density of $n_{e,0}\sim0.08$ cm$^{-3}$, and a fairly consistent minimum density of $\sim$0.003 cm$^{-3}$. Assuming average core temperatures of $\sim$5 keV, these maxima and minima correspond to central cooling times of 0.5\,Gyr and 11.2\,Gyr, respectively.
The lack of evolution in the distribution of central densities (and, by extension, cooling times) suggests that the fraction of cool cores, and the properties of these cores, is relatively stable over the redshift range covered \citep[see also][]{vikhlinin07,santos08,santos10,mcdonald13b}. If there were a higher or lower fraction of cool/non-cool cores at high-$z$ than at low-$z$, we would expect this to manifest in the measured averages.

We note that, while we attempted to mask point sources, there may be contributions to the surface brightness (and gas density) profile from undetected point sources. Assuming a realistic source density, these will have a negligible effect at large radii, but could bias the density high in the innermost bins. This is an issue that we can not address with the available data, but we do note that all of the trends reported here are the same whether we consider the central density or the second radial bin, suggesting that X-ray bright central AGN are not driving our results.

\begin{figure}[h!]
\centering
\includegraphics[width=0.45\textwidth]{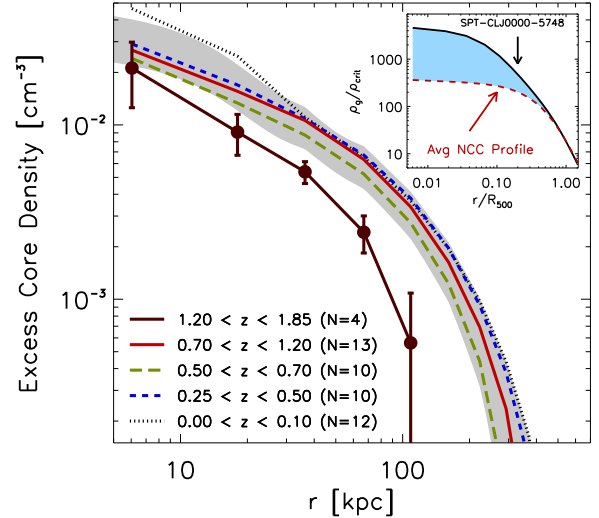}
\caption{Mean over-density profile for cool cores as a function of redshift. For each cool core cluster ($n_{e,0} > 1.5\times10^{-2}$ cm$^{-3}$), we subtract the average non-cool core profile (based on 33 clusters), as shown in the inset in the upper right. The shaded blue region represents the residual overdensity as a function of radius for this one cluster. In each redshift bin, we average these overdensity profiles, yielding the curves shown in the larger panel. The grey region represents the mean and 1$\sigma$ scatter for the full sample of cool cores. This figure demonstrates that the normalization and size of cool cores has not evolved in a significant way since $z\sim1.2$, with a hint ($\sim$2$\sigma$, based on only 4 clusters) of evolution in the highest redshift bin.
\vskip -0.1in}
\label{fig:cc_profiles}
\end{figure}

\begin{figure*}[htb]
\centering
\includegraphics[width=0.995\textwidth]{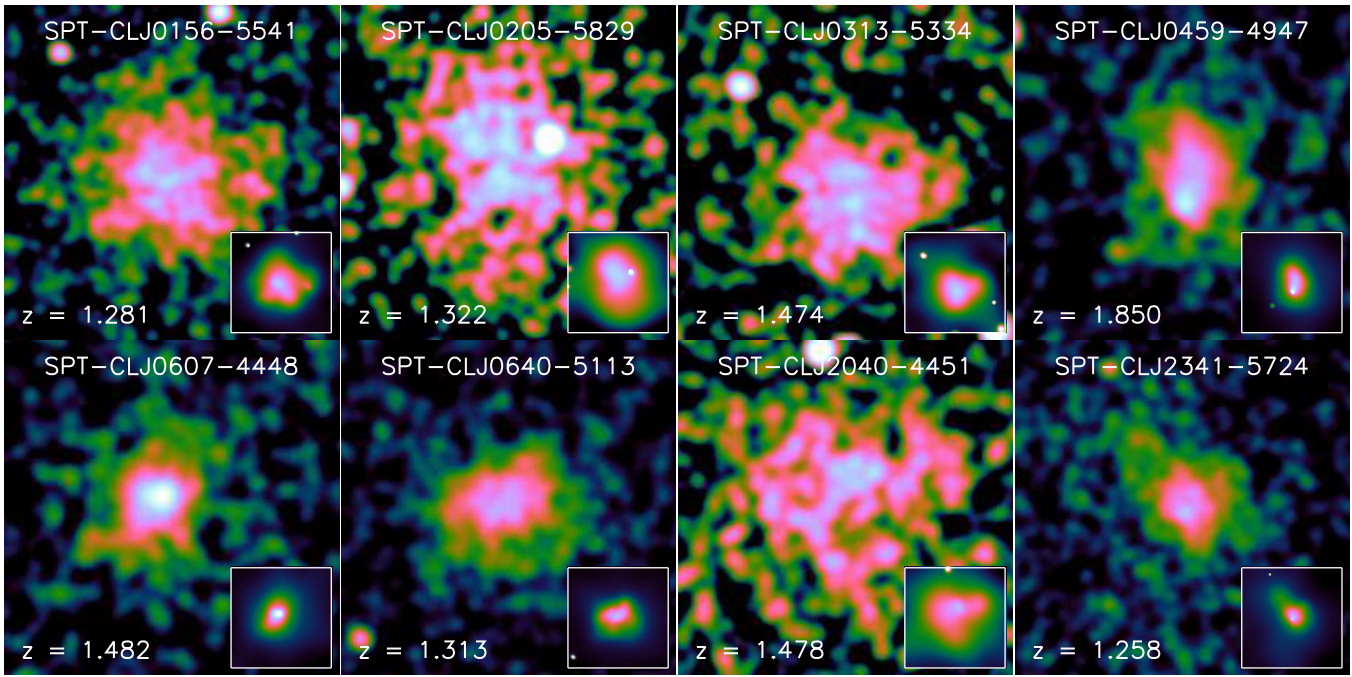}
\caption{0.5--4.0 keV X-ray images of the 8 clusters in the SPT-Hi$z$ sample. Each image spans 3$\times$R$_{500}$ on a side, and has been smoothed with a fixed-width Gaussian with $\textsc{fwhm} = 5^{\prime\prime}$. In the insets, we show adaptively-smoothed images, where the smoothing conditions have been chosen to suppress noise and highlight real structure. As discussed in \S3.1, this smoothing has been tested on low-$z$, high signal-to-noise data to ensure that noise peaks are not being identified as real structures. This figure shows the diversity of X-ray morphologies for the 8 clusters in our sample.}
\label{fig:images_zgt12}
\end{figure*}

We next consider the \emph{shape} of cool cores as a function of redshift. To determine the radial cool core profile, we subtract the average non-cool core profile from each cool core cluster, and stack the residuals. This procedure is shown for a single cluster in the inset of Figure \ref{fig:cc_profiles}. Here, we define non-cool cores and cool cores as having $n_{e,0} < 0.5\times10^{-2}$ cm$^{-3}$ and $n_{e,0} > 1.5\times10^{-2}$ cm$^{-3}$, respectively, avoiding the ``moderate cool core'' regime \citep[see e.g.,][]{hudson10}. Each of these divisions (cool core, moderate cool core, non-cool core) contain roughly a third of the cluster sample. The average cool core profile, derived from 49 cool core clusters spanning $0 < z < 1.9$, is shown as the shaded region in Figure \ref{fig:cc_profiles}, and is well fit by a $\beta$-model with a core radius of $\sim$20--30 kpc. Integrating this profile yields a total cool core gas mass of $\sim3.5\times10^{12}$ M$_{\odot}$, compared to a median total gas mass for these clusters of $5.5\times10^{13}$ M$_{\odot}$.

When we divide the cool core sample into redshift slices, we find no evolution in the shape of the cool core. Within the uncertainties, the four residual profiles, spanning $z=0$ to $z=1.2$, lie on top of each other. The only exception to this is the highest-redshift bin, where the core appears to be considerably smaller in radius. We caution that this result is at the $\sim$2$\sigma$ level, and is based on only 4 cool core clusters identified at $z>1.2$. It is nonetheless intriguing, and may be an indication that we are approaching the epoch of cool core formation at $z\sim1.6$.

The combination of Figures \ref{fig:nepeak} and \ref{fig:cc_profiles} demonstrate that the fraction of clusters harboring cool cores, the central density of cool cores, and the size/shape of cool cores have not evolved significantly in the past $\sim$9 Gyr ($z\lesssim1.2$). The fact that cool cores are confined to the inner $\sim$100\,kpc at all redshifts is consistent with the idea that, on large scales, cool core and non-cool cores are indistinguishable \citep[e.g.,][]{medezinski16}.
The data hint at an epoch of core formation at $z>1.2$, but with only 8 clusters at such high redshifts, this result is not statistically significant.

\subsection{X-ray Morphology}

The X-ray morphology of a galaxy cluster is commonly used as a probe of the cluster's dynamical state \citep[e.g.,][]{mohr95,schuecker01,weissmann13,mantz15}. Nurgaliev \etal (2016) demonstrated that the measured value of $a_{phot}$, which we use in this work to quantify morphology, is significantly elevated during a major merger for $\sim$1--2 Gyr, based on hydrodynamic simulations of 26 major (M$_1$/M$_2$ $>$ 0.5) cluster mergers. This implies that the redshift evolution of $a_{phot}$ ought to roughly probe the evolution of the merger rate over the redshift range considered here. Before providing quantitative results, however, we consider the X-ray images themselves in an attempt to draw qualitative conclusions on the morphological evolution of massive clusters.

In Figure \ref{fig:images_zgt12}, we show Gaussian smoothed and adaptively smoothed (using \textsc{csmooth}\footnote{\url{http://cxc.harvard.edu/ciao/ahelp/csmooth.html}})  0.5--4.0 keV images of the 8 clusters in our high-$z$ sample. The adaptive smoothing parameters were chosen to highlight substructure, while avoiding the identification of noise peaks as significant. The latter condition was tested on dozens of images of the Bullet and El Gordo clusters, subsampled to 2000 counts each, to determine the appropriate \textsc{csmooth} parameter settings to maximize resolution while minimizing false detections of substructure. This figure demonstrates that the X-ray morphologies of these high-$z$ clusters are not dramatically different than their low-$z$ counterparts. We see evidence for highly-disturbed (elongated) systems (e.g., SPT-CLJ2040-4451, SPT-CLJ2341-5724), systems with cores offset from their centroid which are likely sloshing (e.g., SPT-CLJ0459-4947, SPT-CLJ0205-5829), and relatively relaxed systems (e.g., SPT-CLJ0607-4448, SPT-CLJ0640-5113). We find no obvious major mergers (i.e., two distinct, highly-separated peaks). With the limited signal to noise of these exposures, there is no obvious qualitative bias in the morphology of these clusters when compared to the lower-$z$ systems in the full SPT-XVP sample (Nurgaliev \etal 2016).

\begin{figure}[htb]
\centering
\includegraphics[width=0.49\textwidth]{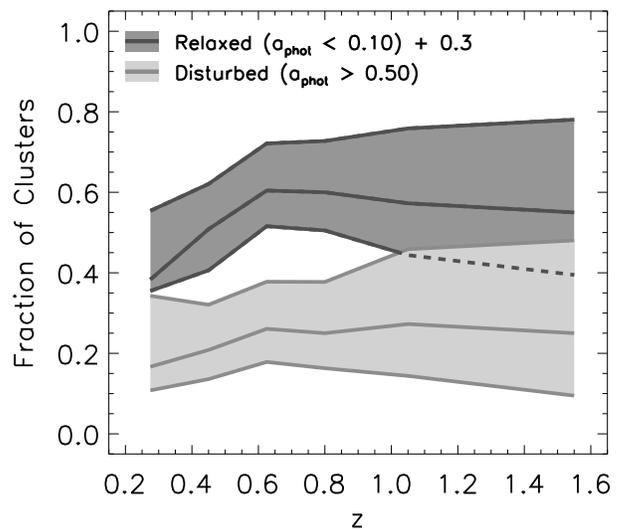}
\caption{Disturbed (light gray) and relaxed (dark gray) fractions as a function of redshift for the SPT-XVP and SPT-Hi$z$ samples, as derived from the X-ray morphology. These fractions are calculated in six independent redshift bins ($z=0.2-0.35,~0.35-0.55, ~0.55-0.7, ~0.7-0.9, ~0.9-1.2, ~1.2-1.9$).
The relaxed fraction has been offset high by 0.3, to allow a more straightforward visual comparison. We have chosen to show only the extremes of the morphological distribution here, excluding all clusters near the relaxed/disturbed boundary. The choice of threshold $a_{phot}$ values for classification as disturbed or relaxed is arbitrary, and does not drive the result. We find that there is no strong evolution in the fraction of clusters with symmetric or highly-asymmetric X-ray morphologies.}
\label{fig:aphotfrac_vsz}
\end{figure}

We consider the dependence of the morphologically disturbed and relaxed fractions as a function of redshift in Figure \ref{fig:aphotfrac_vsz}. In this figure, we arbitrarily define ``relaxed'' as having $a_{phot} < 0.1$ and ``disturbed'' as having $a_{phot} > 0.5$. The latter is somewhat motivated by simulations \citep{nurgaliev16}, and is approximately representative of major (nearly equal mass) mergers. We note that the choice of threshold does not drive our result. The results of Figure \ref{fig:aphotfrac_vsz} are somewhat surprising: we see no significant evolution in the disturbed or relaxed fraction over the full redshift range studied here. This is consistent with what was found by Nurgaliev \etal (2016) for an SPT-selected sample spanning a smaller redshift range, and is seemingly at odds with the increasing merger rate with redshift predicted by simulations \citep[e.g.,][]{fakhouri10b}. The implication of this result is that, over the past $\sim$10 Gyr, there has been no measurable increase in the frequency of major mergers in the most massive clusters. This would either imply that these halos assemble rapidly at $z\gtrsim2$, followed by a slow growth fueled primarily by minor mergers, or that we are missing an important piece of the puzzle. 

Overall, we find no obvious difference in X-ray morphology between our low-$z$ ($0.25 < z < 1.2$) and high-$z$ ($1.2 < z < 1.9$) cluster samples. We will discuss possible reasons for this lack of evolution in \S4.1. We note that, given the relatively low signal-to-noise ratio of these data compared to well-studied low-redshift clusters, we can not make any claims on the evolution of more subtle substructure such as core sloshing, cold fronts, or shocks -- such features require significantly deeper observations to identify.

\section{Discussion}

\subsection{ICM Density Profiles: Comparison to Simulations}

\begin{figure}[htb]
\centering
\includegraphics[width=0.5\textwidth]{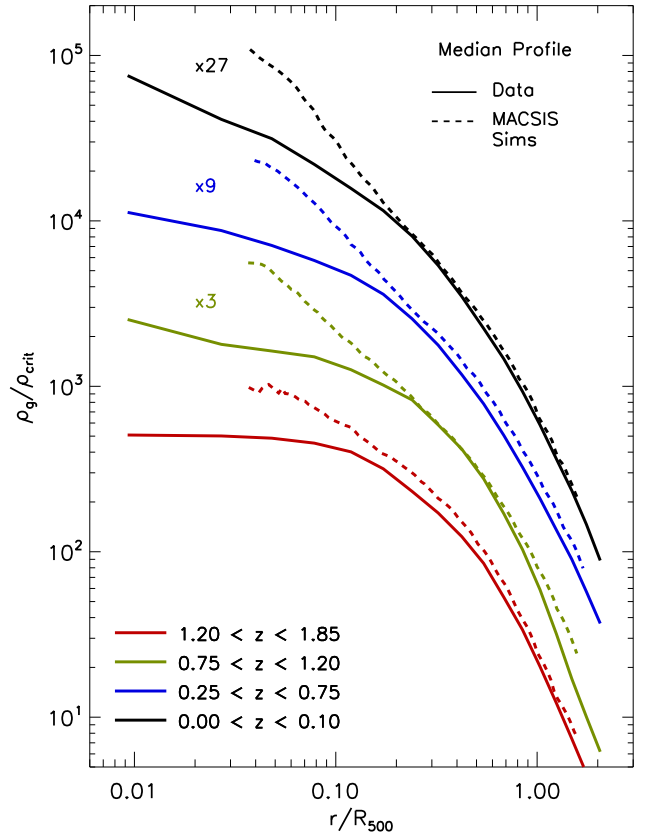}
\caption{Median gas density profiles for observed clusters in four different redshift ranges (solid lines). Profiles have been scaled by arbitrary factors (1, 3, 9, 27) to improve clarity. We also show, with dotted lines, clusters from the MACSIS simulations \citep{barnes17} that have been matched in redshift and mass to the observed systems. At large radii, there is excellent agreement between data and simulations. At small radii ($\lesssim0.1R_{500}$), the simulated clusters are factors of $\sim$2--3 times more dense than their observed counterparts. This disagreement is most likely due to complex interactions between the radio jets in the central AGN and the cooling ICM which are not being fully captured by the simulations.}
\label{fig:macsis}
\end{figure}

\begin{figure*}[htb]
\centering
\includegraphics[width=0.99\textwidth]{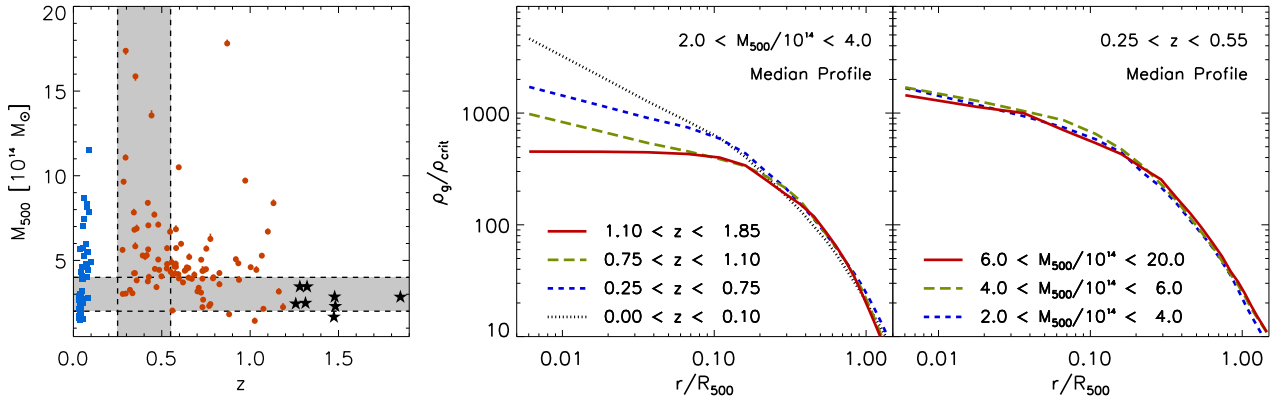}
\caption{\emph{Left Panel:} Distribution of galaxy cluster masses and redshifts used in this work. For the low-$z$ subsample here, drawn from \cite{vikhlinin09a}, we consider a broader mass range than in the previous plots. Grey shaded regions represent cuts for two subsamples: a large mass range at nearly fixed redshift, and a large redshift range at nearly fixed mass. \emph{Center Panel:} Median density profiles for clusters over a broad redshift range and narrow mass range. This shows the same evolution as in Figure \ref{fig:rhog}, suggesting that this was not a result of a mass bias between redshift bins. \emph{Right panel:} Median density profiles for clusters over a broad mass range and narrow redshift range. These median profiles are indistinguishable, suggesting that there is no mass dependence driving our results in Figure \ref{fig:rhog}.}
\label{fig:rhog_fixMz}
\end{figure*}

In \cite{mcdonald14c}, we compared the average pressure profiles of clusters from $z\sim0$ to $z\sim1$ to the latest simulations at the time. Here, we compare the measured density profiles over a larger redshift range to the more recent MACSIS simulations \citep{barnes17}. These simulations track 390 clusters over a large range in cosmic time and mass, including approximations of various baryonic physics processes. Clusters are identified in a large (3.2 Gpc) volume dark matter only simulation with mass resolution of $5.43 \times 10^{10}$ M$_{\odot}/h$ and softening length of 40 kpc, and then re-simulated with hydrodynamics at improved resolution with mass resolution of $4.4 \times 10^9$ M$_{\odot}/h$ and softening length of 3 kpc. For details of these simulations, see \cite{barnes17}. From this sample of simulated clusters, we select subsamples at mean redshifts of $\left<z\right>=0.0,0.5,1.0,1.5$ and with median masses matching those of the observed clusters at each redshift. In Figure \ref{fig:macsis} we show the median gas density profiles, normalized to the critical density, in these four redshift bins for both the observed and simulated clusters. In general, the simulated and observed clusters appear similar at $r>0.2R_{500}$, suggesting that the large scale physics is being properly captured in these simulations. We find offsets of $\sim$10\% in normalization between the real and simulated clusters, which may be due to a number of small differences, including the mean mass per particle (in converting from electron density to mass density), the distribution of masses \citep[low-mass clusters will scatter low in $\rho_g/\rho_{crit}$;][]{vikhlinin06a}, the cluster gas fractions, or the cosmology assumed. These offsets are small, and signify that the physics of the ICM is well-described by simulations outside of cluster cores.
In the cores ($r<0.1R_{500}$), simulated clusters have a factor of $\sim$2--3 higher density than observed clusters at the same redshift, suggesting that the included physics may be insufficient to describe the complex interplay between the central radio-loud AGN, its host giant elliptical galaxy, and the dense cluster core. This is similar to what was reported in \cite{mcdonald14c}, when comparing to simulations from \cite{battaglia12} and \cite{bocquet16}, and is a long-standing problem with creating realistic clusters in cosmological simulations \citep[for a review, see][]{kravtsov12}. This issue appears to be present at all epochs, with clusters at $1.2 < z < 1.9$ having over-dense cores in simulations compared to observations at the same redshift. Within the uncertainty, we measure no significant improvement in the data--simulation comparison in cluster cores over the full redshift range probed here.

In summary, we find that the latest MACSIS simulations \citep{barnes17} yield a good match to the observed density profiles of clusters in this work, at $r>0.2R_{500}$. In cluster cores, the simulations over-predict the ICM density by a factor of $\sim$2--3 at all redshifts.

\subsection{Understanding the Evolution of Cluster Cores}

In \S3.1, we showed that the inner slope of the median gas density profile has evolved significantly over the past $\sim$10 Gyr. We first investigate whether this is due to mass evolution in our sample, by isolating first a narrow range in mass and considering the redshift dependence and then isolating a narrow range in redshift and considering the mass dependence. For this test, we include lower-mass systems from \cite{vikhlinin09a}, for a direct (non-evolving) comparison to the low-mass systems at $z>1.2$.
In Figure \ref{fig:rhog_fixMz} we show the results of this test, where we have used coarser redshift bins than in Figure \ref{fig:rhog} since the number of clusters in the narrow mass range is small. We find that, even in a very narrow mass range ($14.3 < \log_{10}M_{500} < 14.6$), there is a strong redshift dependence, with the low-$z$ clusters having significantly cuspier density profiles than their high-$z$ counterparts. In contrast, if we consider an order of magnitude range in mass at roughly fixed redshift ($0.25 < z < 0.55$), we measure no significant variation in the median gas density profile. This suggests that the core evolution shown in Figure \ref{fig:rhog}, and reported in \cite{mcdonald13b}, is not a byproduct of the mass evolution of clusters, but is indeed a steady change in the median density slope over the past $\sim$10 Gyr for clusters at a fixed mass.

Figures \ref{fig:rhog}, \ref{fig:selfsim}, \ref{fig:nepeak}, and \ref{fig:cc_profiles} reveal several important features about the ICM density profiles in massive clusters. Namely, we find remarkable similarity in the \emph{absolute} properties of cool cores as a function of redshift, including the distribution of core densities, the average central density, and the shape of the cool core excess density profile. The lack of observable evolution in any of these properties suggests that the three-dimensional shape and quasi thermal equilibrium of cool cores were established early in the evolution of clusters. These properties have been maintained over timescales significantly longer than the cool core cooling time, suggesting that the source of feedback that is offsetting cooling is tightly self-regulated. Such a tight loop between the cool core properties and the feedback response can be achieved via chaotic cold accretion, i.e., cold clouds and filaments condense out of the hot ICM and are efficiently funneled toward the black hole via inelastic collisions \citep[e.g.,][]{gaspari16,tremblay16}, triggering the immediate AGN outflow response and thus preventing the catastrophic steepening of density profiles.

At the same time, we find no evidence for departures from self similar evolution at radii larger than 0.2R$_{500}$. Interestingly, this is precisely the radius at which the average temperature profile for cool core clusters deviates from that of non-cool core clusters \citep{vikhlinin06a,baldi12}. We conclude that, to within the precision of our measurements, the ICM density profile has evolved self similarly at $r>0.2$R$_{500}$ over the past $\sim$10\,Gyr.

\begin{figure}[htb]
\centering
\includegraphics[width=0.45\textwidth]{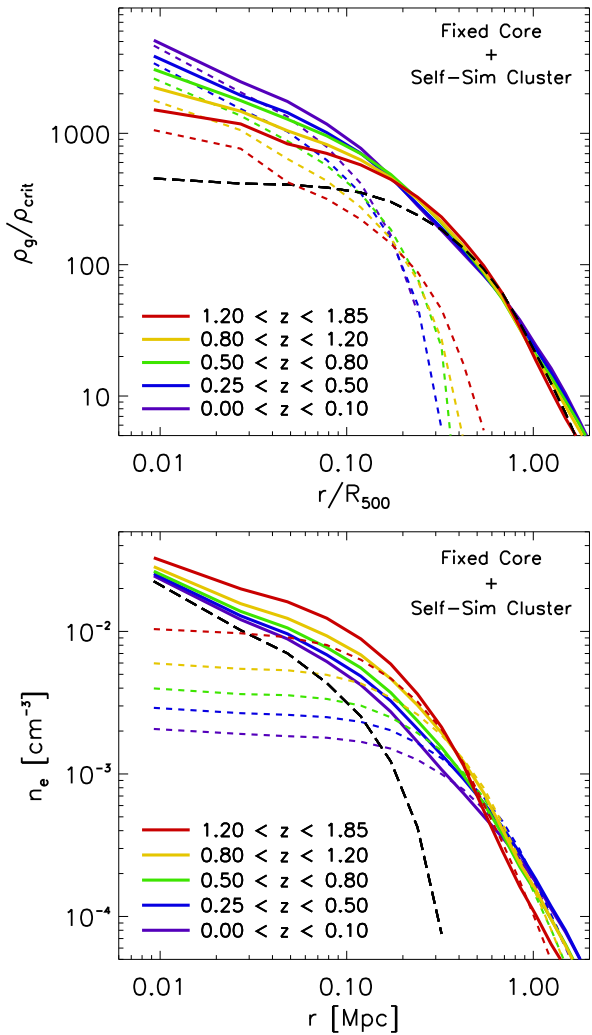}
\caption{\emph{Upper panel:} Expected density profiles (solid lines) for a self similarly-evolving, non-cool core cluster (dashed black line) combined with a non-evolving cool core (dotted colored lines). Because of the choice of scaling, the non-evolving cool core term appears to be evolving. \emph{Lower panel:} Same as above, but now showing the profiles in absolute physical units. Without any cosmological scaling, the cool core now appears nearly static while the bulk of the cluster shows the expected self similar evolution.}
\label{fig:cc_growth}
\end{figure}

The above two paragraphs describe a scenario in which the properties of cool cores are locked in early, while the rest of the cluster evolves in a predictable fashion that is well-described by simple models of gravitational collapse \citep[see e.g.,][]{kravtsov12}. This two-stage evolution is demonstrated in Figure \ref{fig:cc_growth}. This figure shows that the evolution in cuspiness that we see in Figure \ref{fig:rhog}, and that was previously reported by \cite{vikhlinin07}, \cite{santos08,santos10}, and \cite{mcdonald13b}, can be reproduced with a non-evolving core embedded in a self similarly-`evolving cluster. The evolving cuspiness, in this scenario, is due to the increasing contrast between the dense cool core and the rest of the cluster which, at high-$z$, is at higher density for a given r/R$_{500}$.

This result appears to, at first, contradict the evolving core mass presented in \cite{mcdonald13b}. In this previous work, the mass of the cool core was defined as the difference between the cool core and average non-cool core profile (as defined here), but only integrated to 0.1R$_{500}$. Because R$_{500}$ is a physically smaller radius for high-$z$ clusters, this meant that we were integrating over much less of the core volume for high-$z$ clusters than for their low-$z$ counterparts. Since the cool core does not appear to be evolving in size, it makes more sense to define the outer radius in physical units (i.e., 100\,kpc) rather than relative units (i.e., 0.1R$_{500}$).

In summary, we find that the evolution in the ICM density profiles for massive clusters from $z=0$ to $z\sim1.6$ is well-described by the sum of a self similarly-evolving non-cool core profile and a non-evolving cool core. This simple picture describes the results presented here (Figures \ref{fig:rhog}, \ref{fig:selfsim}, \ref{fig:nepeak}, and \ref{fig:cc_profiles}) and in previous works \citep[e.g.,][]{vikhlinin07,santos08,santos10,mcdonald13b}. The size of the unevolving core, approximately 100-200\,kpc, provides a rough boundary within which the similarity-breaking feedback mechanism (i.e., AGN feedback) must do work. The fact that the core has remained stable in size and mass over such a long time period indicates that AGN feedback must be tightly regulated and gentle, instead of being injected via a strong quasar blast \citep{gaspari14}.

\subsection{The Evolution of the Halo Merger Rate}

\begin{figure}[t]
\centering
\includegraphics[width=0.5\textwidth]{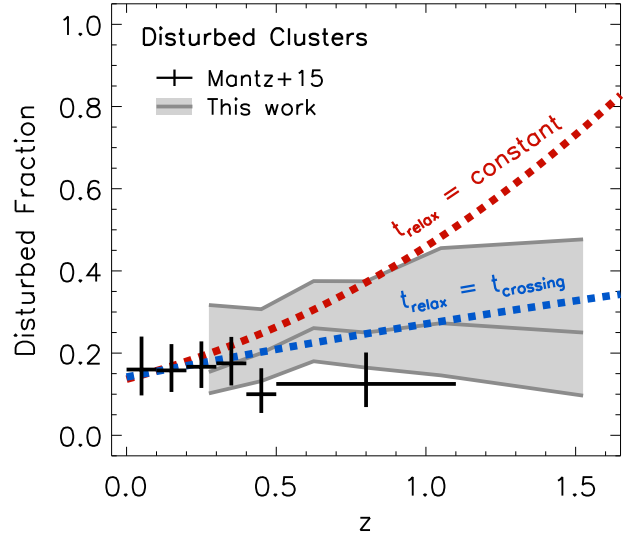}
\caption{This figure shows the fraction of observed clusters morphologically classified as ``disturbed'' as a function of redshift, from Figure \ref{fig:aphotfrac_vsz}. We also include data from \cite{mantz15}, where the disturbed fraction is defined based on their ``symmetry'' parameter, which we find agrees well with $a_{phot}$ for identifying disturbed systems. We have excluded SPT-selected clusters from the \cite{mantz15} study for this comparison. We compare these data to the halo merger rate for massive halos from \cite{fakhouri10}, assuming that a cluster appears disturbed after a major merger for a fixed amount of time (red line) or for a crossing time (blue line), and normalizing the profiles to agree with the data at $z\sim0.1$. The latter agrees well with the data, and implies that clusters at early times relaxed faster after a merger than those today, due to their lower mass and higher density.
}
\label{fig:aphotfrac_pred}
\end{figure}

In Figure \ref{fig:aphotfrac_vsz}, we showed that the fraction of clusters identified as ``disturbed'' based on asymmetry in the X-ray emission has not changed significantly from $z\sim0.2$ to $z\sim1.4$. This appears to contradict the prediction from simulations that the merger rate is a strong function of redshift \citep[see e.g.,][]{fakhouri10}, but is consistent with other groups that have studied the evolution of cluster morphology \citep[e.g.,][]{nurgaliev16,mantz15}. For the most massive halos ($M>10^{14}$ M$_{\odot}$), \cite{fakhouri10} find that the rate of major ($M_1/M_2 > 0.3$) mergers, $dN_m/dt$, increases from $\sim$0.07 Gyr$^{-1}$ at $z\sim0$ to 0.2 Gyr$^{-1}$ at $z\sim1$, or roughly a factor of 3 increase over the past $\sim$8 Gyr. 
However, to go from a predicted halo merger rate to an observed disturbed fraction, we must assume a timescale over which the X-ray emission would appear disturbed after a major merger (the ``relaxation time''). The simplest choice of relaxation time would be one that is constant with redshift, meaning that the observed disturbed fraction would trace the halo merger rate. Figure \ref{fig:aphotfrac_pred} shows how poorly this choice of timescale fares, when compared to the data from both this work and from \cite{mantz15}\footnote{We use the ``symmetry'' ($S$) parameter from \cite{mantz15} to identify disturbed clusters. Using overlapping clusters from the analyses of \cite{mantz15} and \cite{nurgaliev16}, we find that $S<0.6$ is roughly equivalent to $a_{phot} >0.5$.} . At the highest redshifts probed, the predicted evolution is inconsistent with the observations at the $>$97\% confidence level, suggesting that the choice of a constant relaxation time is a poor one.

However, if we modify our assumption about how long a cluster will appear disturbed in the X-rays after a major merger, we predict a dramatically different evolution. Assuming self similar growth of clusters, the crossing time  \citep[$\tau_{cr} \propto R/\sigma \propto H(z)^{-1}$;][]{carlberg97} ought to be shorter at early times. Under the assumption that a cluster appears disturbed for approximately a crossing time (or, relaxation time is proportional to crossing time), the expected disturbed fraction from simulations is highly suppressed. This is due to the fact that the merger rate is $\sim$3 times higher at $z\sim1$ compared to $z\sim0$ \citep{fakhouri10}, while the relaxation time is $\sim$2 times shorter ($H(z)$ is twice as large) over the same redshift interval. Combined, this results in a relatively mild evolution, fully consistent with what is observed (Figure \ref{fig:aphotfrac_pred}).

\subsection{Demographics of Massive, High-$z$ Clusters}

\begin{figure}[h!]
\centering
\includegraphics[width=0.49\textwidth]{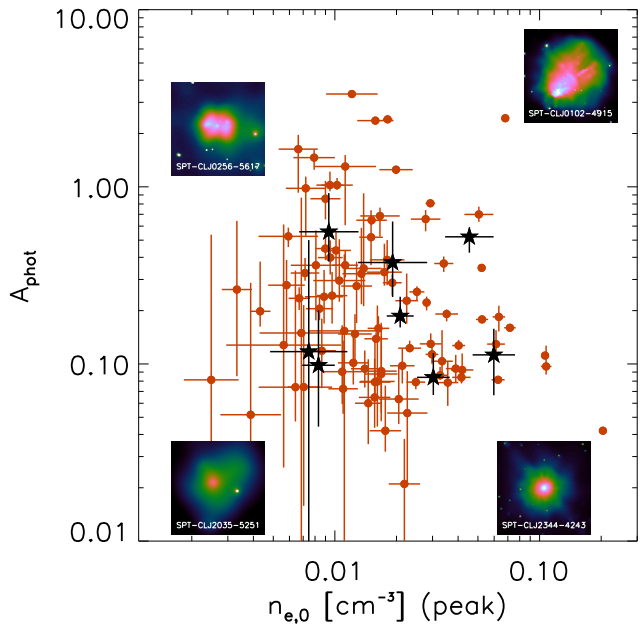}
\caption{Photon asymmetry (A$_{phot}$) versus peak density ($n_{e,0}$) for the clusters in the SPT-XVP (red circles) and SPT-Hi$z$ (black stars) samples. We show small X-ray surface brightness maps for four low-$z$ clusters in the extreme corners of this plot, demonstrating disturbed and relaxed clusters with and without density peaks. The 8 high-$z$ clusters span the full range of morphologies, occupying all parts of this parameter space.}
\label{fig:aphot_ne0}
\end{figure}

Using the combination of the peak density and the morphological asymmetry, we can consider cluster morphologies in two dimensions: radial and azimuthal. In Figure \ref{fig:aphot_ne0}, we show the distribution of clusters in the SPT-XVP and SPT-Hi$z$ samples in this two dimensional space, which roughly separates clusters into four categories: relaxed cool cores, disturbed cool cores, relaxed non-cool cores, and disturbed non-cool cores. We find that the high-$z$ clusters occupy the full range of parameter space, with each of the four types represented clearly in this sample of 8 clusters. Interestingly, one of the 10 strongest cool cores in the full sample is at $z\sim1.5$, suggesting that cool cores were able to form very early on. Overall, we see no evidence that a specific morphological class is over- or under-represented in this $z>1.2$ sample.

\section{Summary}

We have presented results from an X-ray study of 8 SZ-selected galaxy clusters at $z>1.2$ and M$_{500} > 2\times10^{14}$ M$_{\odot}$, which were observed recently with the \emph{Chandra X-ray Observatory}. We combine this sample of high-$z$ clusters with samples of 49 massive X-ray selected clusters at $0 < z < 0.1$, and 90 SZ-selected clusters spanning $0.25 < z < 1.2$, all with existing \emph{Chandra} data, allowing us to track the evolution of the ICM over $\sim$10 Gyr. In this work, we focus specifically on quantities derived based on the X-ray surface brightness, and defer a spectroscopic analysis to a future paper. Below, we summarize the main results of this study.

\begin{itemize}

\item We find that, at $r>0.2R_{500}$, the ICM density profiles of massive galaxy clusters are fully consistent with expectations from self similar evolution (i.e., $n_e \propto E(z)^2$), over the full redshift range probed here. At $r<0.2R_{500}$, we find departures from self similarity, with the centers of clusters showing no significant evolution in gas density ($n_{e,0.01R_{500}} \propto E(z)^{0.1 \pm 0.5}$).

\item Consistent with earlier works, we find that the central ``cuspiness'' of ICM density profiles continues to decrease with increasing redshift, while the absolute central density remains constant, on average.

\item We find that the mean over-density profile of cool cores does not evolve, with the central density, radial extent, and total integrated mass remaining constant from $z=0$ to $z=1.2$. There is a ($\sim$2$\sigma$) hint of evolution at $z>1.2$, based on only 4 cool core clusters at these high redshifts.

\item We propose an evolutionary scenario in which cool cores formed early ($z\gtrsim1.5$) and their properties (size, mass, density) have remained fixed, while the bulk of the cluster has grown in size and mass around them. The combination of a fixed core and a self similarly-evolving cluster provides a successful description of our observations, and suggests that AGN feedback, mainly affecting the inner $\sim$100 kpc scale, is preserving the core properties for over $\sim$10 Gyr in a gentle and tightly self-regulated way.

\item We find that clusters at $z>1.2$ span the same range in morphology as those at $z<0.5$, with no measurable bias towards an overabundance of relaxed or merging systems. This sample of 8 systems includes one that we would classify as a relaxed, strong cool core, and two that we would classify as being highly disturbed.

\item We confirm and extend previous works by \cite{nurgaliev16} and \cite{mantz15}, who show that there is no measurable evolution in the fraction of clusters morphologically classified as ``disturbed'' (i.e., major mergers). We show that this is consistent with the rapidly rising merger rate predicted by cosmological simulations, if we assume that the relaxation time scales like the crossing time (which, on average, decreases with increasing redshift).

\end{itemize}

In summary, we find that the properties of the most distant clusters observed with \emph{Chandra} are remarkably similar to the well-studied systems at $z\sim0$. The cores of clusters appear to be ``frozen'' in time, the bulk of the cluster is evolving self similarly, and the fraction of relaxed/disturbed clusters has not changed significantly. Given the fact that high redshift clusters are both faint and redshifted to low energy, where current X-ray telescopes are less sensitive, it will be challenging to significantly improve upon the constraints provided here. The combination of future cluster surveys, including those in the SZ such as SPT-3G \citep{benson14} and Advanced ACT-Pol \citep{niemack10}, and infrared (e.g., WFIRST, Euclid), coupled with a next-generation X-ray telescopes (e.g., Star-X, Athena, Lynx), will provide orders of magnitude improvement on analyses such as this one, and allow us to trace the properties of the ICM back to its appearance at $z\sim2-3$.


\section*{Acknowledgements} 
Much of this work was enabled by generous GTO contributions from Stephen S. Murray, the Chandra High Resolution Camera PI.  The work was in progress at the time of his untimely death in 2015. He was a valued member of the Center for Astrophysics and a strong supporter of SPT science - he will be greatly missed by all of us.
Support for this work was provided by the National Aeronautics and Space Administration through Chandra Award Number GO5-16141X issued by the Chandra X-ray Observatory Center, which is operated by the Smithsonian Astrophysical Observatory for and on behalf of the National Aeronautics Space Administration under contract NAS8-03060.
The South Pole Telescope is supported by the National Science Foundation through grant PLR-1248097. Partial support is also provided by the NSF Physics Frontier Center grant PHY-1125897 to the Kavli Institute of Cosmological Physics at the University of Chicago, the Kavli Foundation and the Gordon and Betty Moore Foundation grant GBMF 947.
BB is supported by the Fermi Research Alliance, LLC under Contract No. De-AC02-07CH11359 with the United States Department of Energy.
Work at Argonne National Laboratory was supported under U.S. Department of Energy contract DE-AC02-06CH11357.
MG is supported by NASA through Einstein Postdoctoral Fellowship Award Number PF-160137 issued by the Chandra X-ray Observatory Center, which is operated by the SAO for and on behalf of NASA under contract NAS8-03060.
%


%

\begin{appendix}
%
%
Below, we list the mean (Table \ref{table:meanprofs}) and median (Table \ref{table:medianprofs}) density profiles, both in normalized and absolute units. These data are plotted in Figure \ref{fig:rhog}. Uncertainties quoted are 1$\sigma$ uncertainties on the mean/median.

\setcounter{table}{0}
\renewcommand\thetable{A.\arabic{table}}

\begin{table}[h!]
\caption{Mean ICM Density Profiles}

{\tiny
\centering
\begin{tabular}{c c c c c c | c c c c c c}
\hline\hline
 & & & $z$ & & & & & & $z$\\
  & $0.0-0.1$ & $0.25-0.5$ & $0.5-0.75$ & $0.75-1.2$ & $1.2-1.9$ & & $0.0-0.1$ & $0.25-0.5$ & $0.5-0.75$ & $0.75-1.2$ & $1.2-1.9$ \\
$r/R_{500}$ & & \multicolumn{3}{c}{$\log_{10}$($\rho_g/\rho_{crit}$)} & & $r$ [kpc] & & \multicolumn{3}{c}{$\log_{10}$($n_e$ [cm$^{-3}$])}\\
\hline
0.01 & 3.47 $\pm$ 0.11 & 3.20 $\pm$ 0.07 & 3.09 $\pm$ 0.07 & 2.96 $\pm$ 0.08 & 2.84 $\pm$ 0.15 &    6 & -1.82 $\pm$  0.12 & -1.99 $\pm$  0.07 & -1.99 $\pm$  0.07 & -1.97 $\pm$  0.08 & -1.87 $\pm$  0.13\\
0.02 & 3.26 $\pm$ 0.09 & 3.09 $\pm$ 0.06 & 2.98 $\pm$ 0.06 & 2.86 $\pm$ 0.06 & 2.76 $\pm$ 0.12 &   18 & -2.04 $\pm$  0.10 & -2.09 $\pm$  0.06 & -2.10 $\pm$  0.06 & -2.07 $\pm$  0.06 & -1.96 $\pm$  0.10\\
0.04 & 3.11 $\pm$ 0.07 & 3.01 $\pm$ 0.05 & 2.90 $\pm$ 0.05 & 2.80 $\pm$ 0.05 & 2.70 $\pm$ 0.09 &   36 & -2.19 $\pm$  0.08 & -2.17 $\pm$  0.05 & -2.18 $\pm$  0.05 & -2.13 $\pm$  0.06 & -2.02 $\pm$  0.08\\
0.07 & 2.95 $\pm$ 0.06 & 2.92 $\pm$ 0.04 & 2.81 $\pm$ 0.04 & 2.73 $\pm$ 0.05 & 2.64 $\pm$ 0.07 &   67 & -2.33 $\pm$  0.07 & -2.26 $\pm$  0.04 & -2.27 $\pm$  0.04 & -2.21 $\pm$  0.05 & -2.11 $\pm$  0.05\\
0.11 & 2.80 $\pm$ 0.05 & 2.80 $\pm$ 0.03 & 2.71 $\pm$ 0.03 & 2.65 $\pm$ 0.04 & 2.57 $\pm$ 0.05 &  109 & -2.47 $\pm$  0.06 & -2.37 $\pm$  0.03 & -2.38 $\pm$  0.03 & -2.30 $\pm$  0.04 & -2.23 $\pm$  0.04\\
0.16 & 2.64 $\pm$ 0.04 & 2.67 $\pm$ 0.02 & 2.60 $\pm$ 0.02 & 2.56 $\pm$ 0.03 & 2.49 $\pm$ 0.04 &  161 & -2.61 $\pm$  0.05 & -2.49 $\pm$  0.03 & -2.51 $\pm$  0.02 & -2.41 $\pm$  0.04 & -2.36 $\pm$  0.03\\
0.22 & 2.48 $\pm$ 0.03 & 2.52 $\pm$ 0.02 & 2.47 $\pm$ 0.01 & 2.46 $\pm$ 0.03 & 2.40 $\pm$ 0.03 &  222 & -2.75 $\pm$  0.04 & -2.63 $\pm$  0.03 & -2.64 $\pm$  0.01 & -2.54 $\pm$  0.04 & -2.52 $\pm$  0.03\\
0.29 & 2.33 $\pm$ 0.02 & 2.37 $\pm$ 0.01 & 2.34 $\pm$ 0.01 & 2.34 $\pm$ 0.02 & 2.29 $\pm$ 0.02 &  292 & -2.89 $\pm$  0.03 & -2.77 $\pm$  0.03 & -2.78 $\pm$  0.01 & -2.68 $\pm$  0.04 & -2.70 $\pm$  0.03\\
0.37 & 2.19 $\pm$ 0.01 & 2.22 $\pm$ 0.01 & 2.20 $\pm$ 0.01 & 2.22 $\pm$ 0.02 & 2.18 $\pm$ 0.02 &  370 & -3.02 $\pm$  0.02 & -2.91 $\pm$  0.03 & -2.93 $\pm$  0.01 & -2.83 $\pm$  0.04 & -2.89 $\pm$  0.04\\
0.45 & 2.05 $\pm$ 0.01 & 2.08 $\pm$ 0.01 & 2.07 $\pm$ 0.00 & 2.09 $\pm$ 0.01 & 2.05 $\pm$ 0.02 &  454 & -3.15 $\pm$  0.02 & -3.04 $\pm$  0.03 & -3.07 $\pm$  0.01 & -2.99 $\pm$  0.04 & -3.07 $\pm$  0.04\\
0.54 & 1.92 $\pm$ 0.01 & 1.94 $\pm$ 0.01 & 1.94 $\pm$ 0.00 & 1.95 $\pm$ 0.01 & 1.92 $\pm$ 0.02 &  544 & -3.27 $\pm$  0.02 & -3.18 $\pm$  0.03 & -3.21 $\pm$  0.02 & -3.15 $\pm$  0.04 & -3.26 $\pm$  0.05\\
0.64 & 1.79 $\pm$ 0.01 & 1.81 $\pm$ 0.01 & 1.81 $\pm$ 0.01 & 1.82 $\pm$ 0.01 & 1.79 $\pm$ 0.01 &  638 & -3.38 $\pm$  0.02 & -3.30 $\pm$  0.03 & -3.35 $\pm$  0.02 & -3.31 $\pm$  0.05 & -3.44 $\pm$  0.07\\
0.74 & 1.67 $\pm$ 0.01 & 1.69 $\pm$ 0.01 & 1.68 $\pm$ 0.01 & 1.68 $\pm$ 0.01 & 1.66 $\pm$ 0.01 &  737 & -3.50 $\pm$  0.02 & -3.42 $\pm$  0.03 & -3.48 $\pm$  0.02 & -3.47 $\pm$  0.05 & -3.60 $\pm$  0.08\\
0.84 & 1.55 $\pm$ 0.01 & 1.57 $\pm$ 0.01 & 1.56 $\pm$ 0.01 & 1.55 $\pm$ 0.01 & 1.54 $\pm$ 0.02 &  838 & -3.60 $\pm$  0.02 & -3.53 $\pm$  0.03 & -3.61 $\pm$  0.03 & -3.62 $\pm$  0.05 & -3.76 $\pm$  0.09\\
0.94 & 1.44 $\pm$ 0.01 & 1.47 $\pm$ 0.01 & 1.45 $\pm$ 0.01 & 1.42 $\pm$ 0.01 & 1.41 $\pm$ 0.03 &  941 & -3.70 $\pm$  0.02 & -3.64 $\pm$  0.03 & -3.72 $\pm$  0.03 & -3.76 $\pm$  0.05 & -3.90 $\pm$  0.10\\
1.04 & 1.34 $\pm$ 0.01 & 1.36 $\pm$ 0.01 & 1.34 $\pm$ 0.02 & 1.29 $\pm$ 0.02 & 1.30 $\pm$ 0.04 & 1045 & -3.80 $\pm$  0.02 & -3.73 $\pm$  0.03 & -3.83 $\pm$  0.03 & -3.89 $\pm$  0.06 & -4.03 $\pm$  0.12\\
1.15 & 1.24 $\pm$ 0.02 & 1.27 $\pm$ 0.02 & 1.24 $\pm$ 0.02 & 1.18 $\pm$ 0.02 & 1.19 $\pm$ 0.05 & 1149 & -3.89 $\pm$  0.02 & -3.83 $\pm$  0.03 & -3.94 $\pm$  0.04 & -4.02 $\pm$  0.06 & -4.16 $\pm$  0.13\\
1.25 & 1.14 $\pm$ 0.02 & 1.18 $\pm$ 0.02 & 1.15 $\pm$ 0.02 & 1.07 $\pm$ 0.03 & 1.09 $\pm$ 0.06 & 1252 & -3.98 $\pm$  0.02 & -3.91 $\pm$  0.04 & -4.03 $\pm$  0.04 & -4.13 $\pm$  0.06 & -4.27 $\pm$  0.14\\
1.35 & 1.05 $\pm$ 0.02 & 1.10 $\pm$ 0.02 & 1.06 $\pm$ 0.02 & 0.97 $\pm$ 0.03 & 1.00 $\pm$ 0.07 & 1353 & -4.06 $\pm$  0.02 & -3.99 $\pm$  0.04 & -4.12 $\pm$  0.04 & -4.24 $\pm$  0.07 & -4.37 $\pm$  0.15\\
1.45 & 0.97 $\pm$ 0.02 & 1.02 $\pm$ 0.03 & 0.98 $\pm$ 0.03 & 0.88 $\pm$ 0.03 & 0.91 $\pm$ 0.08 & 1452 & -4.14 $\pm$  0.02 & -4.06 $\pm$  0.04 & -4.20 $\pm$  0.04 & -4.33 $\pm$  0.07 & -4.46 $\pm$  0.15\\
\hline
\end{tabular}
\label{table:meanprofs}
}
\caption{Median ICM Density Profiles}
{\tiny
\begin{tabular}{c c c c c c | c c c c c c}
\hline
~~0.01~~ & 3.57 $\pm$ 0.14 & 3.21 $\pm$ 0.08 & 3.07 $\pm$ 0.09 & 2.99 $\pm$ 0.10 & 2.70 $\pm$ 0.19 &    ~~~6~~~ & -1.74 $\pm$  0.15 & -1.99 $\pm$  0.08 & -1.99 $\pm$  0.09 & -2.03 $\pm$  0.09 & -1.91 $\pm$  0.16\\
0.02 & 3.25 $\pm$ 0.11 & 3.09 $\pm$ 0.07 & 2.99 $\pm$ 0.07 & 2.85 $\pm$ 0.08 & 2.70 $\pm$ 0.14 &   18 & -2.07 $\pm$  0.12 & -2.08 $\pm$  0.07 & -2.08 $\pm$  0.07 & -2.10 $\pm$  0.08 & -1.93 $\pm$  0.12\\
0.04 & 3.13 $\pm$ 0.09 & 3.03 $\pm$ 0.06 & 2.90 $\pm$ 0.06 & 2.75 $\pm$ 0.07 & 2.69 $\pm$ 0.11 &   36 & -2.18 $\pm$  0.10 & -2.15 $\pm$  0.06 & -2.19 $\pm$  0.06 & -2.16 $\pm$  0.07 & -2.01 $\pm$  0.09\\
0.07 & 2.96 $\pm$ 0.08 & 2.95 $\pm$ 0.05 & 2.79 $\pm$ 0.05 & 2.71 $\pm$ 0.06 & 2.67 $\pm$ 0.09 &   67 & -2.33 $\pm$  0.08 & -2.22 $\pm$  0.05 & -2.30 $\pm$  0.05 & -2.26 $\pm$  0.06 & -2.11 $\pm$  0.07\\
0.11 & 2.80 $\pm$ 0.06 & 2.83 $\pm$ 0.04 & 2.68 $\pm$ 0.04 & 2.65 $\pm$ 0.05 & 2.62 $\pm$ 0.07 &  109 & -2.49 $\pm$  0.07 & -2.37 $\pm$  0.04 & -2.41 $\pm$  0.04 & -2.32 $\pm$  0.06 & -2.23 $\pm$  0.04\\
0.16 & 2.66 $\pm$ 0.04 & 2.65 $\pm$ 0.03 & 2.58 $\pm$ 0.03 & 2.55 $\pm$ 0.04 & 2.53 $\pm$ 0.05 &  161 & -2.63 $\pm$  0.06 & -2.50 $\pm$  0.04 & -2.51 $\pm$  0.03 & -2.42 $\pm$  0.05 & -2.37 $\pm$  0.03\\
0.22 & 2.52 $\pm$ 0.03 & 2.51 $\pm$ 0.02 & 2.46 $\pm$ 0.02 & 2.46 $\pm$ 0.03 & 2.39 $\pm$ 0.04 &  222 & -2.75 $\pm$  0.05 & -2.65 $\pm$  0.03 & -2.64 $\pm$  0.02 & -2.57 $\pm$  0.05 & -2.53 $\pm$  0.04\\
0.29 & 2.37 $\pm$ 0.02 & 2.37 $\pm$ 0.02 & 2.33 $\pm$ 0.01 & 2.36 $\pm$ 0.03 & 2.28 $\pm$ 0.03 &  292 & -2.88 $\pm$  0.04 & -2.78 $\pm$  0.03 & -2.79 $\pm$  0.01 & -2.68 $\pm$  0.05 & -2.70 $\pm$  0.04\\
0.37 & 2.21 $\pm$ 0.01 & 2.22 $\pm$ 0.01 & 2.19 $\pm$ 0.01 & 2.21 $\pm$ 0.02 & 2.16 $\pm$ 0.03 &  370 & -3.01 $\pm$  0.03 & -2.92 $\pm$  0.03 & -2.93 $\pm$  0.01 & -2.85 $\pm$  0.05 & -2.88 $\pm$  0.05\\
0.45 & 2.06 $\pm$ 0.01 & 2.08 $\pm$ 0.01 & 2.06 $\pm$ 0.01 & 2.09 $\pm$ 0.02 & 2.05 $\pm$ 0.03 &  454 & -3.15 $\pm$  0.03 & -3.07 $\pm$  0.03 & -3.06 $\pm$  0.02 & -2.99 $\pm$  0.05 & -3.06 $\pm$  0.05\\
0.54 & 1.92 $\pm$ 0.01 & 1.94 $\pm$ 0.01 & 1.93 $\pm$ 0.01 & 1.96 $\pm$ 0.01 & 1.93 $\pm$ 0.02 &  544 & -3.28 $\pm$  0.02 & -3.20 $\pm$  0.03 & -3.19 $\pm$  0.02 & -3.15 $\pm$  0.05 & -3.23 $\pm$  0.07\\
0.64 & 1.79 $\pm$ 0.01 & 1.81 $\pm$ 0.01 & 1.81 $\pm$ 0.01 & 1.82 $\pm$ 0.01 & 1.78 $\pm$ 0.01 &  638 & -3.41 $\pm$  0.02 & -3.32 $\pm$  0.03 & -3.32 $\pm$  0.03 & -3.31 $\pm$  0.06 & -3.41 $\pm$  0.08\\
0.74 & 1.67 $\pm$ 0.01 & 1.68 $\pm$ 0.01 & 1.68 $\pm$ 0.01 & 1.68 $\pm$ 0.01 & 1.66 $\pm$ 0.01 &  737 & -3.52 $\pm$  0.02 & -3.46 $\pm$  0.04 & -3.46 $\pm$  0.03 & -3.48 $\pm$  0.06 & -3.59 $\pm$  0.10\\
0.84 & 1.55 $\pm$ 0.01 & 1.57 $\pm$ 0.01 & 1.57 $\pm$ 0.01 & 1.55 $\pm$ 0.01 & 1.54 $\pm$ 0.02 &  838 & -3.62 $\pm$  0.02 & -3.57 $\pm$  0.04 & -3.58 $\pm$  0.03 & -3.59 $\pm$  0.07 & -3.75 $\pm$  0.11\\
0.94 & 1.43 $\pm$ 0.02 & 1.46 $\pm$ 0.02 & 1.47 $\pm$ 0.02 & 1.41 $\pm$ 0.02 & 1.41 $\pm$ 0.04 &  941 & -3.72 $\pm$  0.03 & -3.67 $\pm$  0.04 & -3.69 $\pm$  0.04 & -3.76 $\pm$  0.07 & -3.90 $\pm$  0.13\\
1.04 & 1.32 $\pm$ 0.02 & 1.36 $\pm$ 0.02 & 1.36 $\pm$ 0.02 & 1.28 $\pm$ 0.02 & 1.29 $\pm$ 0.05 & 1045 & -3.82 $\pm$  0.03 & -3.77 $\pm$  0.04 & -3.79 $\pm$  0.04 & -3.92 $\pm$  0.07 & -4.04 $\pm$  0.15\\
1.15 & 1.22 $\pm$ 0.02 & 1.26 $\pm$ 0.02 & 1.26 $\pm$ 0.02 & 1.15 $\pm$ 0.03 & 1.18 $\pm$ 0.06 & 1149 & -3.92 $\pm$  0.03 & -3.85 $\pm$  0.04 & -3.89 $\pm$  0.05 & -4.07 $\pm$  0.08 & -4.17 $\pm$  0.16\\
1.25 & 1.12 $\pm$ 0.02 & 1.17 $\pm$ 0.03 & 1.18 $\pm$ 0.03 & 1.02 $\pm$ 0.03 & 1.07 $\pm$ 0.08 & 1252 & -4.01 $\pm$  0.03 & -3.94 $\pm$  0.04 & -3.98 $\pm$  0.05 & -4.20 $\pm$  0.08 & -4.29 $\pm$  0.17\\
1.35 & 1.04 $\pm$ 0.02 & 1.10 $\pm$ 0.03 & 1.10 $\pm$ 0.03 & 0.90 $\pm$ 0.04 & 0.97 $\pm$ 0.09 & 1353 & -4.09 $\pm$  0.03 & -4.02 $\pm$  0.05 & -4.06 $\pm$  0.05 & -4.33 $\pm$  0.08 & -4.39 $\pm$  0.18\\
1.45 & 0.97 $\pm$ 0.02 & 1.03 $\pm$ 0.03 & 1.03 $\pm$ 0.03 & 0.79 $\pm$ 0.04 & 0.88 $\pm$ 0.10 & 1452 & -4.17 $\pm$  0.03 & -4.09 $\pm$  0.05 & -4.14 $\pm$  0.06 & -4.44 $\pm$  0.08 & -4.49 $\pm$  0.19\\
\hline
\end{tabular}
\label{table:medianprofs}

}
\end{table}

\end{appendix}

\end{document}